\documentclass[10pt,a4paper]{article}
\usepackage[top=2cm,bottom=2cm,left=2cm,right=2cm]{geometry}

\usepackage{amsmath,amssymb}

\usepackage{booktabs}
\usepackage{caption}

\usepackage{changepage}

\usepackage[utf8x]{inputenc}

\usepackage{textcomp,marvosym}

\usepackage{cite}

\usepackage{nameref,hyperref}

\usepackage{microtype}
\DisableLigatures[f]{encoding = *, family = * }

\usepackage[table]{xcolor}

\usepackage{array}

\newcolumntype{+}{!{\vrule width 2pt}}

\newlength\savedwidth

\usepackage{setspace}
\usepackage[aboveskip=1pt,labelfont=bf,labelsep=period,justification=raggedright,singlelinecheck=off]{caption}

\makeatletter
\renewcommand{\@biblabel}[1]{\quad#1.}
\makeatother

\usepackage{algorithm}
\usepackage{algorithmic}
\usepackage{amsthm,amsfonts}
\usepackage{bm}
\usepackage{multirow}

\usepackage{ifthen}

\usepackage{tikz}
\usepackage{tkz-euclide}
\usetkzobj{all}
\usepackage{verbatim}
\usepackage{pgfplotstable}

\newcommand{\M}{\mathcal{M}}
\newcommand{\R}{\mathcal{R}}

\newcommand{\Prob}{\mathbb{P}}
\newcommand{\E}{\mathbb{E}}
\newcommand{\X}{\mathbf{X}}
\newcommand{\y}{\mathbf{y}}

\pgfplotsset{compat=1.14}

\begin{document}
\vspace*{0.2in}

\begin{flushleft}
{\Large
\textbf\newline{Scientific discovery in a model-centric framework: Reproducibility, innovation, and epistemic diversity} 
}
\newline
\\
Berna Devezer\textsuperscript{1,5\Yinyang*},
Luis G. Nardin\textsuperscript{2,5\Yinyang},
Bert Baumgaertner\textsuperscript{3,5\Yinyang},
Erkan Buzbas\textsuperscript{4,5\Yinyang}
\\
\bigskip
\textbf{1} Dept. of Business, University of Idaho, Moscow, ID, USA
\\
\textbf{2} Dept. of Informatics, Brandenburg University of Technology, Cottbus, Germany
\\
\textbf{3} Dept. of Politics and Philosophy, University of Idaho, Moscow, ID, USA
\\
\textbf{4} Dept. of Statistical Science, University of Idaho, Moscow, ID, USA
\\
\textbf{5} Center for Modeling Complex Interactions, University of Idaho, Moscow, ID, USA
\\
\bigskip

%
%
\Yinyang These authors contributed equally to this work.





* bdevezer@uidaho.edu

\end{flushleft}
\section*{Abstract}
Consistent confirmations obtained independently of each other lend credibility to a scientific result. We refer to results satisfying this consistency as reproducible and assume that reproducibility is a desirable property of scientific discovery. Yet seemingly science also progresses despite irreproducible results, indicating that the relationship between reproducibility and other desirable properties of scientific discovery is not well understood. These properties include early discovery of truth, persistence on truth once it is discovered, and time spent on truth in a long-term scientific inquiry. We build a mathematical model of scientific discovery that presents a viable framework to study its desirable properties including reproducibility. In this framework, we assume that scientists adopt a model-centric approach to discover the true model generating data in a stochastic process of scientific discovery. We analyze the properties of this process using Markov chain theory, Monte Carlo methods, and agent-based modeling. We show that the scientific process may not converge to truth even if scientific results are reproducible and that irreproducible results do not necessarily imply untrue results. The proportion of different research strategies represented in the scientific population, scientists' choice of methodology, the complexity of truth, and the strength of signal contribute to this counter-intuitive finding. Important insights include that innovative research speeds up the discovery of scientific truth by facilitating the exploration of model space and epistemic diversity optimizes across desirable properties of scientific discovery.

\section*{Author summary}

We mathematically model a virtual scientific community to understand how reproducibility is related to scientific discovery of truth. Studying this community over time, we find that irreproducible results may not be untrue and reproducible results do not guarantee that the scientific process will converge to truth. A combination of research strategies, statistical methods, and the state of truth explains this finding. We show that innovative research speeds up the discovery of truth and that the pursuit of diverse research approaches in the scientific population optimizes across desirable properties of scientific discovery.

\section*{Introduction}
Consistent confirmations obtained independently of each other lend credibility to a scientific result~\cite{Ramsey1931, Popper1959, Kyburg1964, Schmidt2009}. We refer to this notion of multiple confirmations as \emph{reproducibility of scientific results}. Ioannidis~\cite{Ioannidis2005} argued that a research claim is more likely to be false than true, partly due to the prevalent use of statistical significance and null hypothesis significance testing as method of inference. Recent theoretical research explored aspects of scientific practice contributing to irreproducibility. McElreath and Smaldino~\cite{McElreath2015} modeled a population of scientists testing a variety of hypotheses and tracking positive and negative published findings to investigate how the evidential value of replication studies changed with the base rate of true hypotheses, statistical power, and false positive rate. Other studies found that current incentive structures may lead to degradation of scientific practice~\cite{Smaldino2016, Higginson2016}. Publication bias was also proposed to contribute to the transitioning of incorrect findings from claim to fact~\cite{Nissen2016}. These studies focus on how structural incentives and questionable research practices (QRPs) influence reproducibility of scientific results within a hypothesis-centric framework, and how to improve statistical practices and publication norms to increase reproducibility. Under limitations of hypothesis testing~\cite{Gelman2017}, however, understanding salient properties of the scientific process is challenging, especially for fields that progress by building, comparing, selecting, and re-building models.

In this work, we make three contributions to the literature on meta-research. First, we present a model-centric mathematical framework modeling scientists' convergence to truth in the process of scientific discovery. We identify, mathematically define and study the relationship between key properties of this process such as early discovery of truth, persistence on truth once it is discovered, time spent on truth in a long-term scientific inquiry, and rate of reproducibility. Second, in a system without QRPs or structural incentives, we study how the diversity of research strategies in the scientific population, the complexity of truth, and the noise-to-signal ratio in the true data generating model affect these properties. Third, we study the scientific process where scientists engage in model comparison instead of statistical hypothesis testing. Model comparison aims to select a useful model that approximates the true model generating the data and it has long been a cornerstone in many scientific disciplines because of its generality. Our model-centric view allows us to study the process of scientific discovery under uncertainty, bypassing the complications inherited from hypothesis testing~\cite{Gelman2017}.

\section*{A model-centric meta-scientific framework}\label{sec:matmethods}
We adopt a notion of confirmation of results in idealized experiments and build a mathematical framework of scientific discovery based on this notion.

\textbf{Model, idealized experiment, replication experiment, and reproducibility.}
We let $K$ be the background knowledge on a natural phenomenon of interest, $M$ be a prediction in the form of a probability model parameterized by $\theta \in \Theta$, that is in principle testable using observables, and $D$ be the data generated by the true model. The degree of confirmation of $M$ by $D$ is assessed by $S,$ a fixed and known method. We define $\xi$, an \textit{idealized experiment}, as $(M,\theta,D,S,K)$.

In an idealized experiment $\xi$, the data $D$ confirms the model $M$ if $\Prob(M|D,K)>\Prob(M|K)$, where $\Prob(M|D,K)$ and $\Prob(M|K)$ are probabilities of $M$ after and before observing the data, respectively. By Bayes's Theorem, $\Prob(M|D,K)/\Prob(M|K)$ is proportional to the likelihood $\Prob(D|M,K)$. Large $\Prob(D|M,K)$ implies high degree of confirmation of $M$. Complex models, however, have a tendency to inflate $\Prob(D|M,K)$ and hence $\Prob(M|D,K)$. As a measure against overfitting, modern model comparison statistics $S$ are not only based on $P(D|M,K)$ but also penalize the complexity of $M$ to prevent inflating the likelihood under complex models. For several well-known $S$, smaller $S(M)$ means the model $M$ is more favorable in a set of models, and we follow this convention here.

In a scientific inquiry, a novel prediction is often tested against a status quo consensus. We formulate this situation by denoting the novel prediction as a \textit{proposed model} $M_{P}$ which is tested against the \textit{global model} $M_{G}$, the scientific consensus. Conditional on the data, $S(M_P)<S(M_G)$ means that the proposed model is more favorable than $M_{G}$. In this case, $M_P$ becomes the new scientific consensus, otherwise the global model remains as the scientific consensus.

The description of scientific inquiry given in the last paragraph and \textit{reproducibility of results} in a replication experiment as follows. If $\xi_1$ given by $(M_{P},\theta, D_1, S, K_1)$ is tested against $M_G$, then the experiment $\xi_2$ immediately following $\xi_1$ is a replication experiment for $\xi_1$ if and only if $\xi_{2}$ is given by $(M_{P},\theta, D_{2}, S, K_{2})$ and it is tested against the same $M_G$ as $\xi_1.$ That is, the replication experiment proposes the same model, uses the same methods, and is tested against the same global model as the original experiment. Of two elements that differ between the original experiment and the replication experiment, the first is $D_2$, which is the data that is generated in the replication experiment independent of the data $D_1$ of the original experiment. The second is the background information $K_{2}$ which includes all the information necessary to replicate $\xi_1$. In particular, $K_2$ includes the knowledge that $M_P$ was the proposed model in $\xi_1$, it was tested against $M_G$, and the outcome of this test---whether $M_P$ was updated to consensus or $M_G$ remained as the consensus. We say that the replication experiment $\xi_2$ \textit{reproduces the results} of $\xi_1$ if the results of $\xi_1$ and $\xi_2$ are the same in terms of updating the consensus. There are two mutually exclusive ways that $\xi_2$ can reproduce the results of $\xi_1$:
$1)$ If the proposed model in $\xi_1$ won against the global model, then this must also be the case in $\xi_{2}$, that is $S(M_{P})<S(M_G)$ in both experiments. $2)$ If the proposed model in $\xi_1$ lost against the global model, then this must also be the case in $\xi_{2}$, that is $S(M_{P})>S(M_G)$ in both experiments. Otherwise, we say that $\xi_{2}$ fails to reproduce the results of $\xi_{1}$. These definitions of replication experiment and reproducibility of results formalize necessary open science practices for potential reproducibility of results: Information about the proposed and global models in the original experiment and the results of this experiment which we capture by $K_{2}$ must be transferred to $\xi_{2}$.

\begin{figure}[!ht]
\includegraphics[width=1\textwidth]{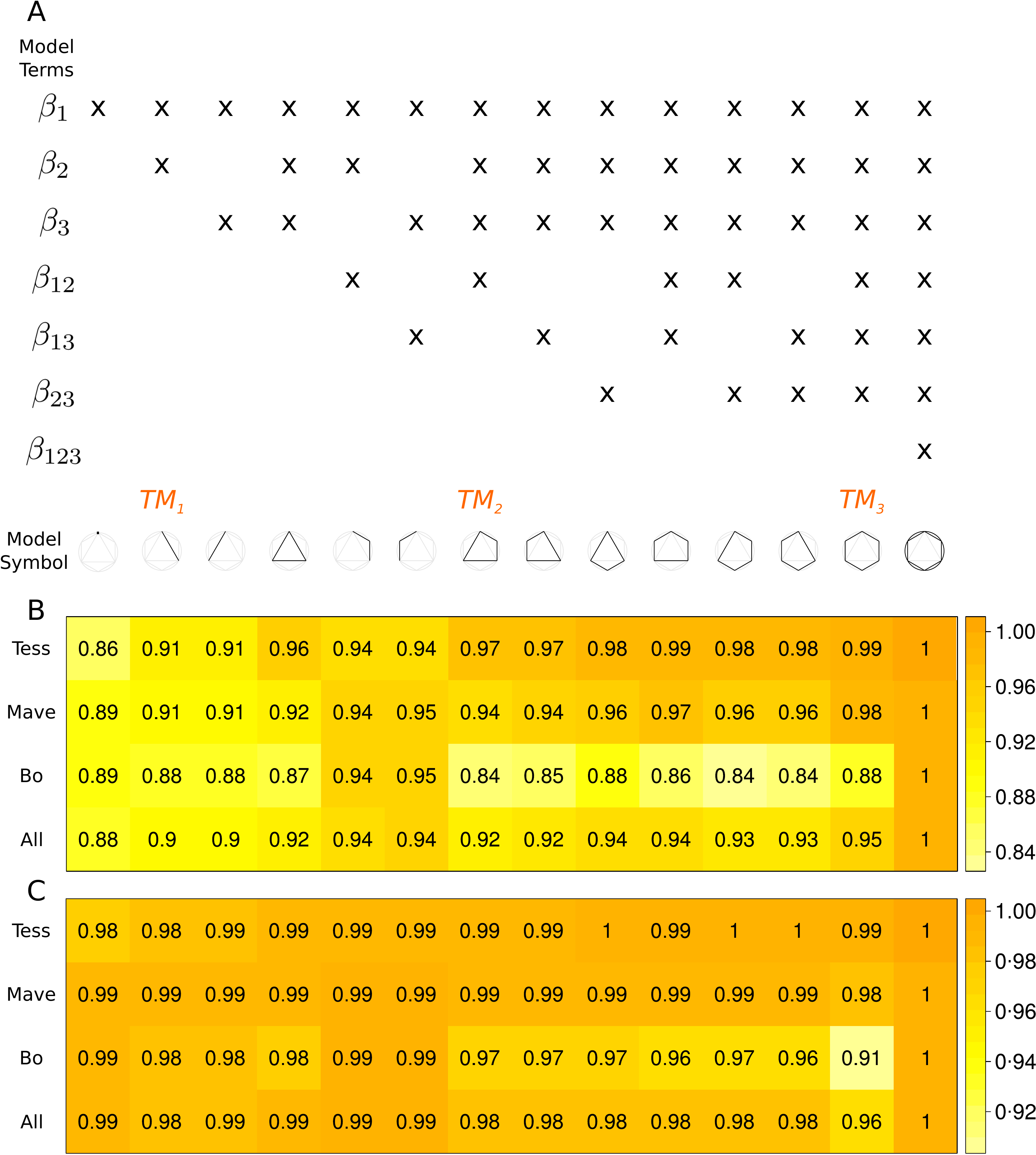}
\caption{(A) Each column of the matrix indicates the terms included in the model shown by a symbol at the bottom of the column. For example, the fifth column denotes the model $y=\beta_1x_1+\beta_2x_2+\beta_{12}x_1x_2+\epsilon$, and is represented by three corners of hexagon connected with two lines. $TM_1, TM_2, TM_3$ are the three true models used in our agent-based model simulations. Symbols representing each model are ordered from simple to complex, left to right. Symbols are used as y-axis labels for heat maps in (B) and (C). Stickiness of each true model as a global model for each scientist population under AIC (B) and SC (C).}
\label{fig:sticky}
\end{figure}

\textbf{Stochastic process of scientific discovery.}
We assume an infinite population of scientists who conduct a sequence of idealized experiments to find the true model generating the data (see \nameref{S1_Appendix} for mathematical framework). In the population, we consider various types of scientists, each with a mathematically well-defined research strategy for proposing models. Scientists search for a true model in a set of linear models. Linear models were chosen because they can accommodate a variety of designs with straightforward statistical analysis, and their complexity is mathematically tractable (\nameref{S2_Appendix}). We define model complexity as a function of the number of model parameters and interaction terms, and visualize it by representing each model with a unique geometry on an equilateral hexagon inscribed in its tangent circle (Fig.~\ref{fig:sticky}A).
%

We assume a discrete time process with $t = 0, 1, \cdots$, where at each time step an experiment $\xi^{(t)}$ is conducted by a scientist randomly selected from a population of scientists with equal probability. The experiment entails proposing a model $M_P^{(t)}$ as a candidate for the true data generating mechanism. The probability of proposing a particular model is determined by the scientist's research strategy and the global model $M_G^{(t)}$---the current scientific consensus. The scientist compares the global model against the proposed model using new data $D^{(t)}$ generated from the true model and a model comparison statistic $S$. The model with favorable statistic is set as the global model for the next time step $M_G^{(t+1)}$. Because the probability of proposing a model is independent of the past and the transition from $M_G^{(t)}$ to $M_G^{(t+1)}$ admits the (first order) Markov property, the stochastic process representing the scientific process is a Markov chain. This mechanism represents how scientific consensus is updated in light of new evidence. We study the mathematical properties of this process for different scientist populations representing a variety of research strategies.

Introducing replication experiments to the process fundamentally alters the probability mechanism of updating global models: By definition, a replication experiment depends on the experiment conducted at the previous time step via $K_2$. Hence the stochastic process is a higher order Markov chain (see~\cite{Bensonetal2017}) and we turn to an agent-based model (ABM)~\cite{Epstein2006, Gilbert2008} to analyze the process with replication. Our ABM is a forward-in-time, simulation-based, individual-level implementation of the scientific process where {\em agents} represent scientists (\nameref{S3_Algorithm}).

We assume that reproducibility is a desirable property of scientific discovery. However, arguably early discovery of truth, persistence on truth once it is discovered, and long time spent on truth in a long-term scientific inquiry are also desirable properties of scientific discovery since they would characterize a resource efficient and epistemically accurate scientific process. We seek insight into the drivers of these properties and the relationship among them, which we motivate with the following questions (see~\nameref{S4_Appendix} for mathematical definitions). \textit{How quickly does scientific community discover the true model?} We assess this property by the mean first passage time to the true model and view it as a key indicator of resource efficiency in the process of scientific discovery. \textit{How ``sticky'' is the true model as global model?} We define the stickiness of the true model as the mean probability of staying in the true model once it becomes global model. \textit{How long does scientific community stay on the true model?} The stationary probability that the true model is global model has the interpretation of the long-term stay of the scientific community on the true model. \textit{How reproducible are the results of experiments?} We track replication experiments to calculate the rate of reproducibility when the true model is global model, as well as when it is not. We study the answers to these questions as a function of the following aspects of model-centric approach to scientific discovery: the proportion of research strategies in the scientific population, the complexity of the true model generating the data, the ratio of error variance to deterministic expectation in the true model (i.e., noise-to-signal ratio), and the model comparison statistic.

\begin{figure}[!ht]
\includegraphics[width=1\textwidth]{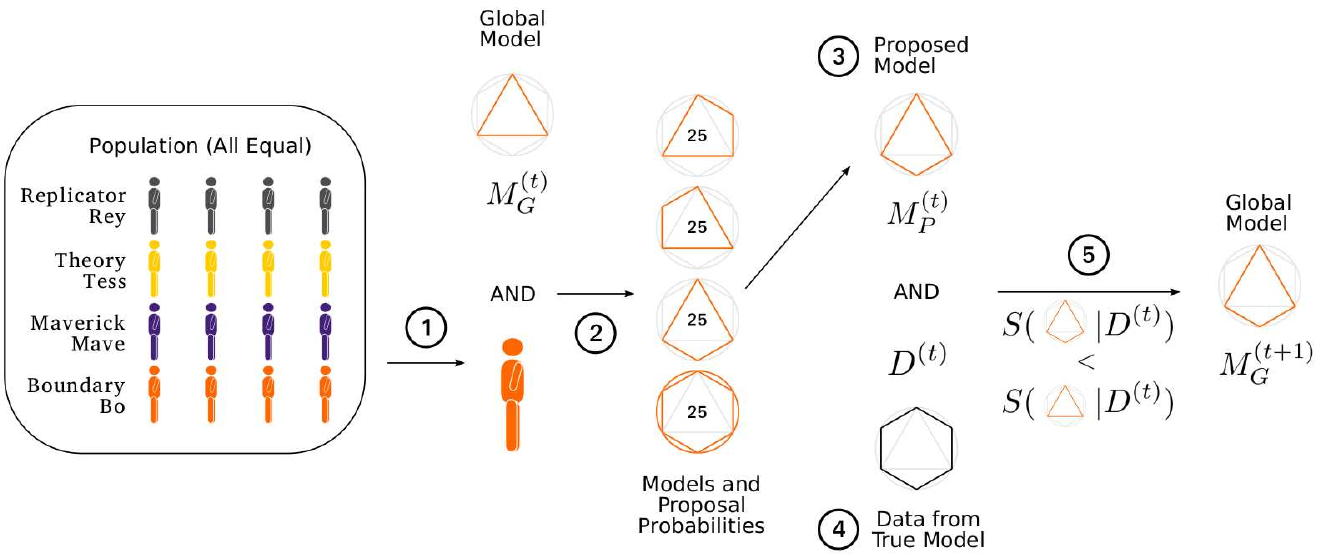}
\caption{A transition of our process of scientific discovery for an epistemically diverse population with replicator. A scientist (Bo) is chosen uniformly randomly from the population (1). Given the global model, the set of proposal models and their probabilities (given in percentage points inside models) are determined. In this population with no replicator, Bo proposes only models formed by adding an interaction (2). The proposed model selected (3) and the data generated from the true model (4) are used with the model comparison statistic (SC or AIC) to update the global model (5).}
\label{fig:example}
\end{figure}

\textbf{Scientist types. }We include distinct research strategies to explore the effect of epistemic diversity in the scientific population. We define simple research strategies where our scientists do not have a memory of their past decisions and they do not interact directly with each other, but only via the global model. Nonetheless, the research strategies we include in our model seem reasonably realistic to us in capturing the essence of some well-known research approaches. Figure~\ref{fig:example} illustrates our stochastic process of scientific discovery for a specific scientist population.

For the process with no replication, we define three types of scientists: $\textit{Tess, Mave, Bo}$ (\nameref{S5_Appendix}). \textit{Tess, the theory tester}, uniformly randomly selects a proposal model that is only one main effect predictor away from the current global model. We impose a hierarchical constraint on interaction terms in the sense that when {\em Tess} proposes to drop a predictor from the current global model, all higher order interactions including this term are dropped too. We think of {\em Tess}' strategy as a refinement of an existing theory by testing current consensus against models that are close to it. {\em Mave, the maverick} does not build off of the current consensus but she ignores it. She advances novel ideas and uniformly randomly selects a model from the set of all models. The novelty-seeking aspect of her strategy is similar to a maverick strategy proposed in prior research on epistemic landscapes and the division of cognitive labor~\cite{Weisberg2009, Muldoon2013, Alexander2015}. However, in contrast to that strategy, {\em Mave} does not actively aim to avoid previously tested models and she acts independently of the current scientific consensus. {\em Bo, the boundary tester} systematically tests the boundaries of the current global model. She selects a model that adds interactions to the current global model to explore the conditions under which the current global model holds. After her, if the tested boundary has been confirmed by the data, the global model may earn a new predictor representing a main interaction and the lower order interactions associated with it that are not currently in the model. If the tested boundary has not been confirmed by the data, the global model does not change.

For the process with replication, we introduce {\em Rey, the replicator} who conducts a replication experiment which is the exact same experiment conducted by her precedent, but with new data. She reproduces the results of the preceding experiment if she confirms as global model the same model confirmed as global model by her precedent. Thus, {\em Rey} compares the same pair of proposed model ($M_{P}$) and global model ($M_{G}$) as that of her predecessor, using new data, and the replication is successful if two conditions are satisfied: 1) either the results of her predecessor and the replication experiment are both judged favorable against the global model ($M_{G}$), or they both are not, leaving scientific consensus unchanged, 2) sufficient information about the the result of her predecessor's experiment is available through the background knowledge of the replication experiment to assess if the first condition holds. This second condition implies that a replication experiment requires open science practices for transferring sufficient information about the experiment in the previous time step to the replication experiment in order for the latter to assess the reproducibility of the result.

\textbf{Scientist populations.} We assess the effect of each strategy by considering populations of scientists in which \textit{Rey}, \textit{Tess}, \textit{Bo}, and \textit{Mave} are represented at varying proportions (\hyperref[S6_Table]{S6 Table}). Of particular interest to us are homogeneous populations where the dominant scientist type comprises 99\% of the scientist population and epistemically diverse populations where all scientist types are represented in equal proportion.

\textbf{Accumulation of evidence.} The background knowledge that a scientist brings to an experiment consist of the global model, all other models in the system, predictors, and their parameters as well as the results from the previous experiment if the current experiment is a replication experiment. Scientific evidence in our model accumulates through experiments and all the evidence is counted at the end of a long run. Data set in each experiment has a weight of one and the sample proportion of experiments which reproduce a result converges to the true value of reproducibility rate by the Law of Large Numbers.

\textbf{Model comparison criteria.} We adopt two well-known likelihood-based criteria for model comparison and show how these interact with the behavior of scientists in the population: the Schwarz Criterion (SC)~\cite{Schwarz1978} and the Akaike's Information Criterion (AIC)~\cite{Akaike1973, Akaike1974}. A smaller value of these statistics indicate a better model performance than a larger value. When the true model generating the data is in the universe of candidate models, SC is statistically consistent and selects the true model with probability $1$ as $n\rightarrow \infty$~(\nameref{S5_Appendix}).

\textbf{Maximum number of factors in the model.} For computational feasibility, we fix the number of factors in the linear model to $3$, which results in $14$ models. Each of these $14$ models refers to its linear structure. In this sense, there are infinitely many probability distributions that can be fully specified within a model. For the system without replication, we analyze all $14$ models as true models. The most complex model has 7 predictors (Fig.~\ref{fig:sticky}A), including three main effects, three $2$--way interactions, and one $3$--way interaction. We fix the sample size to $100$ and calibrate the ratio of the error variance $\sigma^2$ to expected value of the model at the mean value of the predictors $\E(y|\mu_x)$ where $\mu_x=\E(x)$. We fix $\sigma^2:\E(y|\mu_x)$ to $(1:4)$ (and include results for $(1:1), (4:1)$ in~\hyperref[S13_Figure]{S13 Figure}-\hyperref[S22_Figure]{S22 Figure};~\nameref{S5_Appendix}).

\textbf{Design of ABM experiments.} For the system with replication, we use three true models representing a gradient of complexity (Fig.~\ref{fig:sticky}A $TM_1,TM_2,TM_3$). We set up a completely randomized factorial simulation experiment: $3$ true models, $3$ $\sigma^2:\E(y|\mu_x)$ levels at $(1:4),(1:1),(4:1)$, $5$ scientist populations (\hyperref[S6_Table]{S6 Table}), and $2$ model comparison statistics (AIC, SC). We run each experimental condition as an ABM simulation for $11000$ iterations and replicated $100$ times, each using a different random seed. We discard the first $1000$ iterations as burn-in, except when analyzing the mean first passage time to true model. Code and data are given in~\nameref{S7_CodeAndData}.

\section*{A brief discussion of our modeling choices, assumptions, and their implications}\label{sec:synopsis}

Our model-centric framework facilitates investigating the consequences of the process of scientific discovery, including reproducibility, as a system-wide phenomenon. System-wide reproducibility and its relationship to scientific discovery are largely unexplored topics. Navigating through numerous potential variables and parameters to create a realistic system rich in behavior whose outcomes are easily interpretable is challenging. Our model aims to create such a system by making design choices and simplifying assumptions. Among many results that we obtain, we report some intuitive results as sanity checks. These results connect our idealized system to reality. However, we highlight the results that seem counter-intuitive to us because we find them to be interesting patterns warranting further investigation. The implications and limitations of each specific result are discussed in the Results section.

Here, we qualitatively clarify the implications and limitations of our system and emphasize the assumptions which constitute its salient features for our results to hold. We anchor our system firmly against the backdrop of guarantees provided by statistical theory to avoid over-generalization.

\textbf{What statistical theory offers in isolation.}
A well-known statistical inference mode is comparing a set of hypotheses represented as probabilistic models, with the goal of selecting a model. A statistical method selects the model which fits the stochastic observations best according to some well-defined measure. Consider the following three conditions:
\begin{enumerate}
\item There exists a true model generating the observations and it is in the search set.
\item The signal in the observations is detectable.
\item A reliable method whose assumptions are met is used to perform inferences.
\end{enumerate}
If these conditions are met, then the statistical theory guarantees that under repeated testing with independent observations, the true model is selected with highest frequency. This frequency approaches to a constant value determined by conditions (1),(2), and (3). The practical implication of this guarantee is that the results under the true model are reproducible with a constant rate.

We now contemplate on the consequences of violating conditions (1)-(3).
If condition (1) is not met, then the true model cannot be selected. In this case, well-established methods select the model that is closest to the true model and in the set with highest frequency. As we discuss in research strategies below, this is a situation where if results are reproducible they are not true.

Conditions (2) and (3) work in conjunction. A method is reliable with respect to the strength of the signal it is designed to detect. There are a variety of ways to evaluate the reliability of statistical methods. Hypothesis tests use specificity and sensitivity. Modern model selection methods often invoke an information-theoretic measure. Intuitively, we expect a statistical test designed to detect the bias in a coin to perform well even with small sample size if the coin is heavy on heads because the data structure is simple and the signal strong. We would be fortunate, however, if a model selection method can discern between two complex models close to each other with the same sample size. If we violate condition (2) or (3), then we have an unreliable method to detect the strength of the signal. In this case, even if the true model generating the observations is in the set, we might not be able to select it with high frequency due to the mismatch between the performance of the method and the strength of the signal (see also~\cite{Baumgaertner2018} for a discussion of how method choice might affect reproducibility).

When conditions (1)-(3) are met, statistical guarantees hold in the absence of external factors that are not part of the data generating mechanism and the inference process. To quote Lindley~\cite{Lindley2000}: ``Statisticians tend to study problems in isolation, with the result that combinations of statements are not needed, and it is in the combinations that difficulties can arise [...]'' Scientific claims often are accompanied by statistical evidence to support them. However, we doubt that in practice scientific discovery is always based on evidence using statistical methods whose assumptions are satisfied. A variety of external factors such as choices made in theory building, design of experiments, data collection, and analyses might affect system-wide properties in scientific discovery. Our work is motivated to develop intuition on how some of these external factors affect the guarantees made by statistical theory. In particular, we introduce external factors which violate conditions (1)-(3), and produce counter-intuitive results. We explicitly discuss two factors that have major effects on our outcomes next.

\textbf{Research strategies as an external factor and their potential counter-intuitive effect on reproducibility.}
When scientists aim to discover a true model among a large number of candidate models, reducing the search space is critical. Our system introduces one external factor to statistical theory as {\em research strategies} which determine the models to be tested at each step of the scientific process thereby serving as a means to reduce the search space. However, by choosing models to reduce the search space, the research strategies also affect the frequency of testing each model. As a consequence of affecting frequencies of tests, these strategies may alter the results guaranteed by statistical theory in many ways. Results depend on how frequently these strategies are employed by the scientific community and how frequently they propose each model. In this sense, the strategies determine the opportunity given to each model to show its value.

To clarify the effect that strategies can have on reproducibility, we give an extreme example. Consider a search space with only three possible models. We pursue the bizarre research strategy to always test two of these models against each other, neither of which is the true model generating the data. Then:
\begin{enumerate}
\item The true model will never be selected because it is never tested.
\item Between the two models tested, the model that is closer to the true model will be selected with higher frequency than the model further.
\item The result stated in item (2) is reproducible.
\end{enumerate}
This toy example shows that we can follow strategies which produce results that are reproducible but not true. In this work, we further show that counter-intuitive results like this can arise under mild research strategies that modify the search space in subtle ways. However, our results {\em do not mean} that true results are not reproducible. In fact, this is a sanity check that we have in our system: provided the three conditions of the previous section are met, {\em true results are reproducible.}

\textbf{System updating as an external factor and its effect on reproducibility.}
A second external factor that our system introduces is the temporal characteristic of the scientific discovery. Probabilistic uncertainty dictates that one instance of statistical inference cannot be conclusive even if the true model is included in the test set and it produces highly reliable data. Thus, repeated testing through time using independent data sets calls for a temporal stochastic process. A state variable defines this process whose outcome is determined as a function of this state variable with respect to a reasonable measure of success.

The natural state variable in our system is the model selected in each test. We think of this model as a pragmatic consensus of the scientific community at any given time in the process of scientific discovery. When another model is proposed, it is tested against this consensus.

There are difficulties in defining a reasonable success measure for models, however. A pragmatic consensus of the scientific community is presumably a model which withstands testing against other models to some degree. The consensus is expected to survive even if it is not selected, say, a few times. A tally of each model against every other model can be kept introducing a system memory. This tally, can be used as prior evidence in the next testing instance of particular models. Introducing memory into the temporal process is technically easy. The real difficulty is how to choose the success measure. A decision rule about when a model should lose confidence and be replaced by another model is needed. Consider the following example: Consensus model A and model B were compared two times each winning once. In the third comparison, model B wins. Should we abandon model A and make model B consensus? If not, how many more times should model B win against model A before we are willing to replace model A?

We find these questions challenging, but they help illustrate our point. One of several well-known rules from decision theory can be implemented to update the consensus. No matter which rule is chosen, however, it will affect system-wide properties including the reproducibility of results. In this sense, a decision rule is another external factor: Precisely because the rule dictates when a model becomes consensus, it has the power to alter the frequency of statistical results in the process, otherwise obtained in isolation.

Even without the complication of a decision rule, scientific strategies make our system complex. They produce a diverse array of results whose implications we do not fully understand. Hence, we left the complication of model memory out of our system by choosing to update the consensus with the selected model at each test. This corresponds to a memoryless $0-1$ decision rule. We admit that this memoryless property of our model is unsatisfactory and might not reflect a realistic representation of the scientific process. We caution the reader to interpret our results with care on this aspect. On the other hand, we are interested in system-wide and aggregate results of our model through time. That is, we look at the rate of reproducibility and other properties of scientific discovery in a given process by integrating across many independent iterations of tests and systems. Thus, our sanity checks still apply. For example, we expect (and find) the scientific consensus to converge to the true model once it is discovered, and the true model to be sticky and reproducible. While the memorylessness of our system might prevent the scientific process being realistically captured at any given point in the process, on the aggregate we are able to observe certain realistic patterns.

\section*{Results}\label{sec:results}


First, we present results in a system with no replicator where properties of our scientific process can be obtained for all true models in our model space using Markov chain theory and computationally efficient Monte Carlo methods (\nameref{S8_Appendix}). We use this computational advantage to gain insight into process properties and to inform ABM experiments for the system with replication, in which exploring all model space is computationally unfeasible. Second, we present results from these ABM experiments (\hyperref[S9_Table]{S9 Table}).

\subsection*{Results in a system with no replication} \label{sec:ResultsNoRep} We examine stickiness of the true model, time spent at the true model, and mean first passage time to the true model for populations composed of different proportions of \textit{Tess, Mave,} and \textit{Bo}. Our interest is in how the proportion of different research strategies represented in scientist populations influences these desirable properties of scientific discovery. A key feature of the theoretical calculations we present in this section is the implementation of \textit{soft} research strategies where all scientists are allowed to propose a model not consistent with their strategy with a small probability. Technically, this feature guarantees that the transition probability matrix of the Markov chain is well connected. In the system with replicator, we investigate \textit{hard} research strategies---where scientists are allowed to propose only models consistent with their strategy---in addition to soft strategies. We compare results across four scientific populations, two model comparison statistics---Akaike's Information Criterion (AIC) and Schwarz Criterion (SC), and all possible true models in our model space (Fig.~\ref{fig:sticky}A). We fix a low error variance to model expectation ratio ($1:4$).

Stickiness, the probability of the true model staying as global model once it is hit, is high under low error both for AIC (Fig.~\ref{fig:sticky}B) and SC (Fig.~\ref{fig:sticky}C). This result serves as a sanity check for our theoretical model. Once the true model becomes the consensus, it stays as such most of the time. The stickiness of the true model increases with complexity, except for the \textit{Bo-}dominant population. For \textit{Bo-}dominant population, stickiness decreases with complexity, except for the most complex true model which \textit{Bo} cannot overfit.

\begin{figure}[H]
\includegraphics[width=1\textwidth]{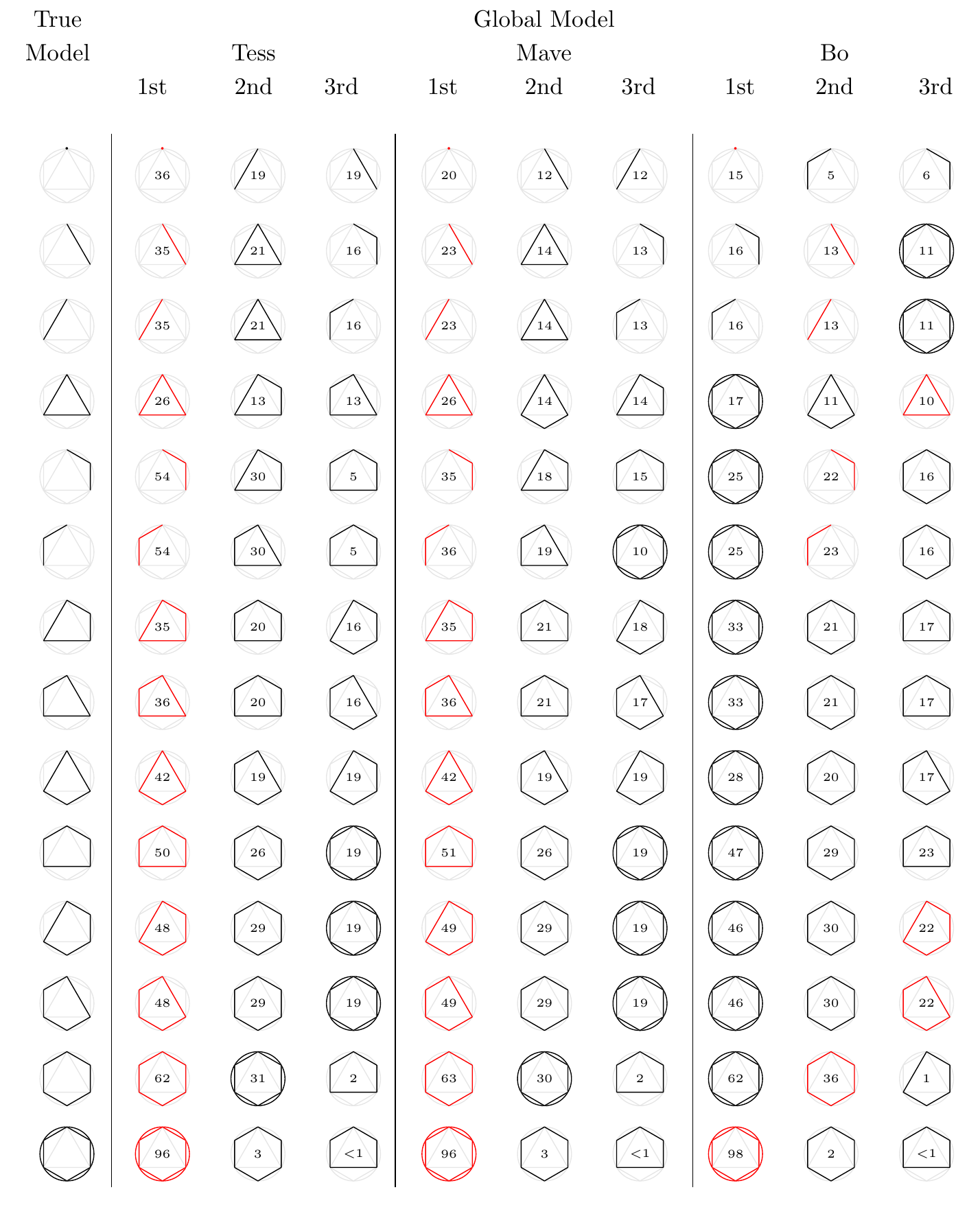}
\captionsetup{labelformat=empty}
\caption{{\bf S10 Figure. The three most visited models for scientist populations with one dominant type and the proportion of time spent at each true model, when AIC is the model comparison statistic and noise-to-signal ratio is $1:4$ in a system with no replication.} For $\sigma^2:\E(y|\mu_x)=(1:4)$, proportion of time spent by a model as the global given a true model, assessed by AIC. For \textit{Tess-}, \textit{Mave-}, and \textit{Bo-}dominant populations, three most visited models are shown. Numbers show time spent at each model in percent points. True models are in red. \textit{Tess-} and \textit{Mave-}dominant populations capture the true model more consistently than \textit{Bo-}dominant populations.}
\label{S10_Figure}
\end{figure}

\begin{figure}[H]
\includegraphics[width=1\textwidth]{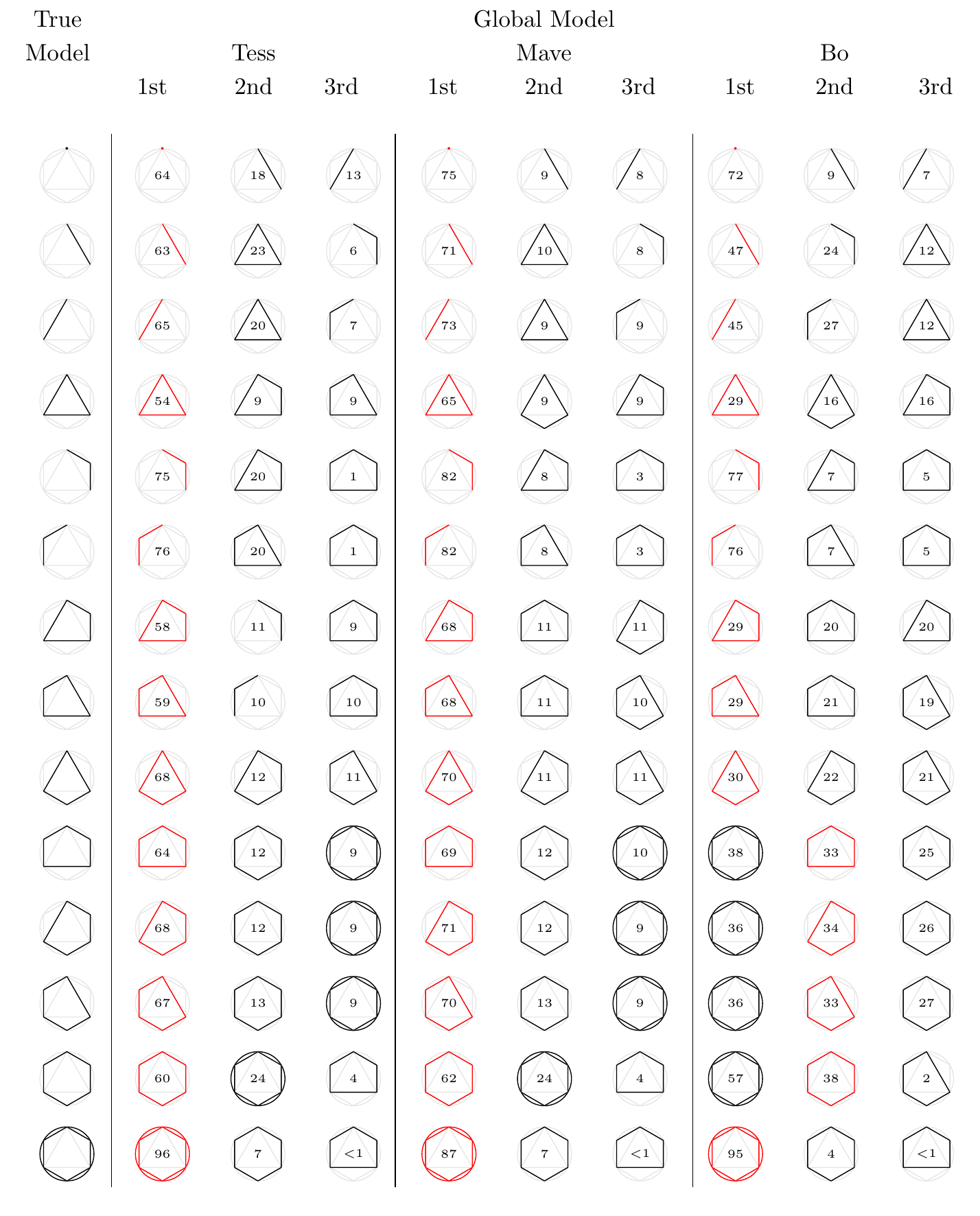}
\captionsetup{labelformat=empty}
\caption{{\bf S11 Figure. The three most visited models for scientist populations with one dominant type and the proportion of time spent at each true model, when SC is the model comparison statistic and noise-to-signal ratio is $1:4$ in a system with no replication.} For $\sigma^2:\E(y|\mu_x)=(1:4)$, proportion of time spent by a model as the global given a true model, assessed by SC. For \textit{Tess-}, \textit{Mave-}, and \textit{Bo-}dominant populations, three most visited models are shown. Numbers show time spent at each model in percent points. True models are in red. \textit{Bo-}dominant population spends much more time at the true model under SC than under AIC.}
\label{S11_Figure}
\end{figure}

Even though the true model is sticky under low error, and hence, tends to stay as global model once it is hit, the system still spends considerable time at models that are not true. For example, \textit{Bo-}dominant population overfits unduly complex models, spending only $25\%$ of the time at the true model under AIC and $48\%$ under SC (\hyperref[S10_Figure]{S10 Figure},~\hyperref[S11_Figure]{S11 Figure}). This population spends most time in models more complex than the true model. Out of $14$ true models in our model space, under AIC $4$ are not among \textit{Bo-}dominant population's top $3$ most visited models, and under SC $4$ are not the most visited model. This is a consequence of \textit{Bo}'s strategy to add only interaction predictors to her proposed models, which regularly pits the global model against more complex models. \textit{Bo}'s poor performance is striking because boundary conditions show whether the relationship between variables holds across the values of other variables and hence, boundary testing is a widely used strategy for theory development in many disciplines~\cite{Mule2017, Whetten1989}. In comparison with \textit{Bo-}dominant population, \textit{Tess-} and \textit{Mave-}dominant populations spend more time in the true model, regardless of its complexity ($47\%$ and $41\%$ under AIC and $67\%$ and $72\%$ under SC, respectively). Overall, the theory testing and maverick strategies maximize the probability of spending time at the true model under AIC and SC, respectively.

\begin{figure}[H]
\includegraphics[width=1\textwidth]{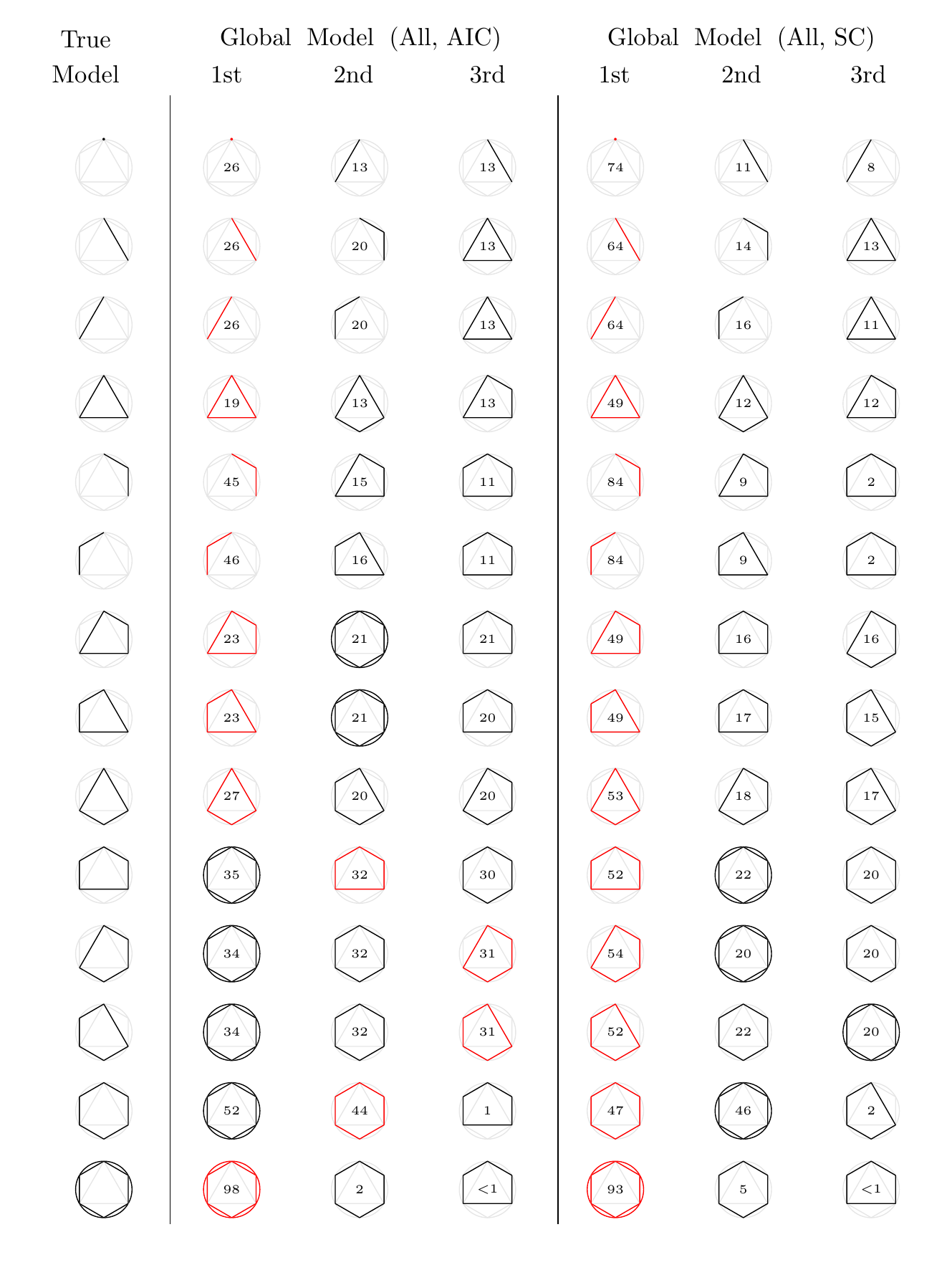}
\captionsetup{labelformat=empty}
\caption{{\bf S12 Figure. The three most visited models by the epistemically diverse population for each true model and when noise-to-signal ratio is $1:4$ in a system with no replication.} For $\sigma^2:\E(y|\mu_x)=(1:4)$, proportion of time spent by a model as the global given a true model for an epistemically diverse population. Three most visited models are shown for AIC and SC. Numbers show time spent at each model in percent points. True models are in red. For both AIC and SC, all true models are in top three most visited models.}
\label{S12_Figure}
\end{figure}

For the epistemically diverse population, the true model is in top $3$ most visited models irrespective of its complexity (\hyperref[S12_Figure]{S12 Figure}). Because boundary testing strategy is ineffective in capturing the true model, the presence of \textit{Bo}s causes the epistemically diverse population to spend less time at the true model ($36\%$ under AIC and $62\%$ under SC) than \textit{Tess-} or \textit{Mave-}dominant populations. However, the effect of overfitting complex models by \textit{Bo} is alleviated in this population due to the presence of other research strategies in the population and does not prevent the epistemically diverse population from consistently recovering the true model. In essence, epistemic diversity protects against ineffective research strategies.

\begin{figure}[!ht]
\includegraphics[width=1\textwidth]{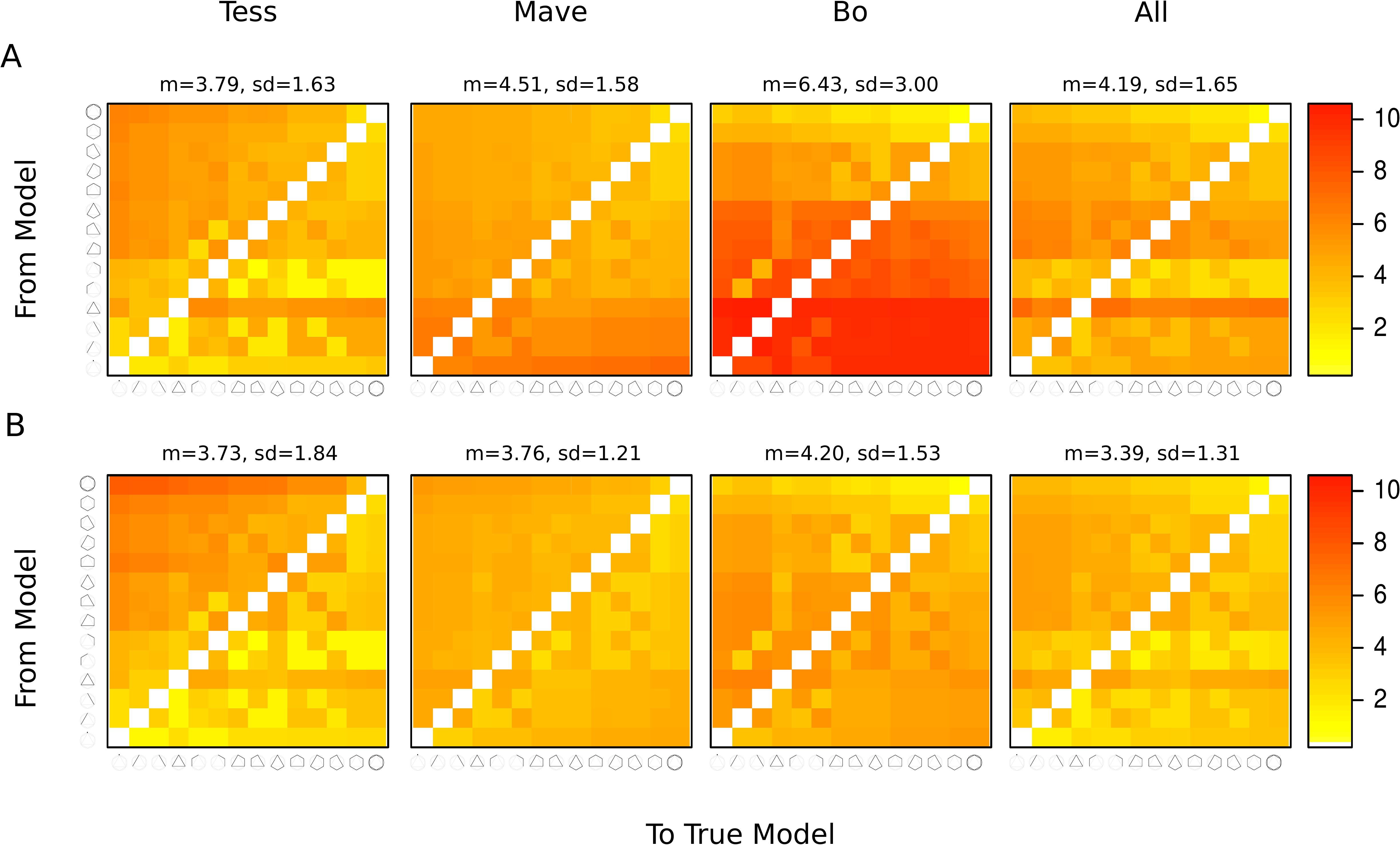}
\caption{The mean first passage time from each initial model (vertical axis) to each true model (horizontal axis) using AIC (A) and SC (B) as model comparison statistics per scientist populations. \textit{All} stands for epistemically diverse; other populations are dominant in the given type.}
\label{fig:MFPT}
\end{figure}

\begin{figure}[H]
\includegraphics[width=1\textwidth]{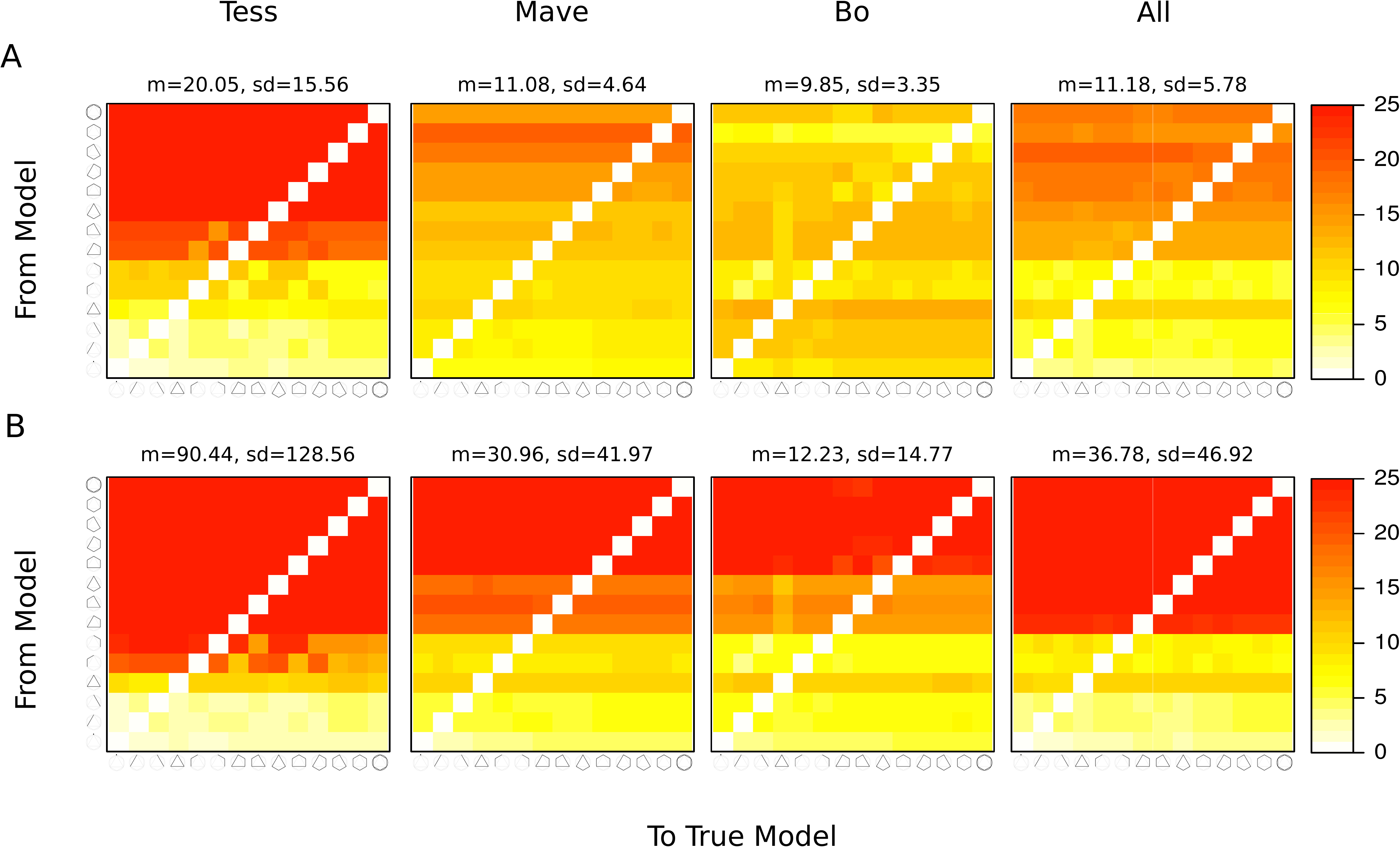}
\captionsetup{labelformat=empty}
\caption{{\bf S22 Figure. The mean first passage time to true model when noise-to-signal ratio is $4:1$ in a system with no replication.} For $\sigma^2:\E(y|\mu_x)=(4:1)$, the mean first passage time from each initial model (vertical axis) to each true model (horizontal axis) using AIC (A) and SC (B) as model comparison statistics per scientist populations. All means epistemically diverse; all others dominant in one type. Due to high variability in this scenario, all values greater than 25 are set to 25 for purposes of illustration. Under high noise, the speed with which the true model is hit is much lower, and the slowest when starting from complex initial models. In this scenario, \textit{Bo}-dominant population is the most efficient out of all four populations.}
\label{S22_Figure}
\end{figure}

We assessed the speed of scientific discovery by the mean first passage time to the true model. In this system without replication, where the transition matrix is well connected due to the implementation of soft research strategies, the true model is hit quickly across all populations (between 3.39 and 6.43 mean number of steps) when noise-to-signal ratio is low. Increasing the proportion of boundary testers in the population, however, slows down the discovery of the true model. Further, the model comparison statistic interacts with scientist populations with respect to the speed of discovery. Under AIC, \textit{Tess-}dominant population is the fastest to find the true model (Fig.~\ref{fig:MFPT}A, \textit{Tess}). In comparison, as shown by red region in Fig.~\ref{fig:MFPT}A, \textit{Bo-}dominant population is slow to discover the true model. \textit{Tess'}s speeding up the discovery of truth is also reflected in the epistemically diverse population. Using SC as opposed to AIC as the model comparison statistic decreases differences across populations (Fig.~\ref{fig:MFPT}B), increasing the speed of discovery for all populations. The fastest population to hit the true model is the epistemically diverse population under SC. We find that the speed of discovery slows down considerably as the noise-to-signal ratio is increased to ($4:1$)
(\hyperref[S22_Figure]{S22 Figure}).

\begin{figure}[H]
\includegraphics[width=1\textwidth]{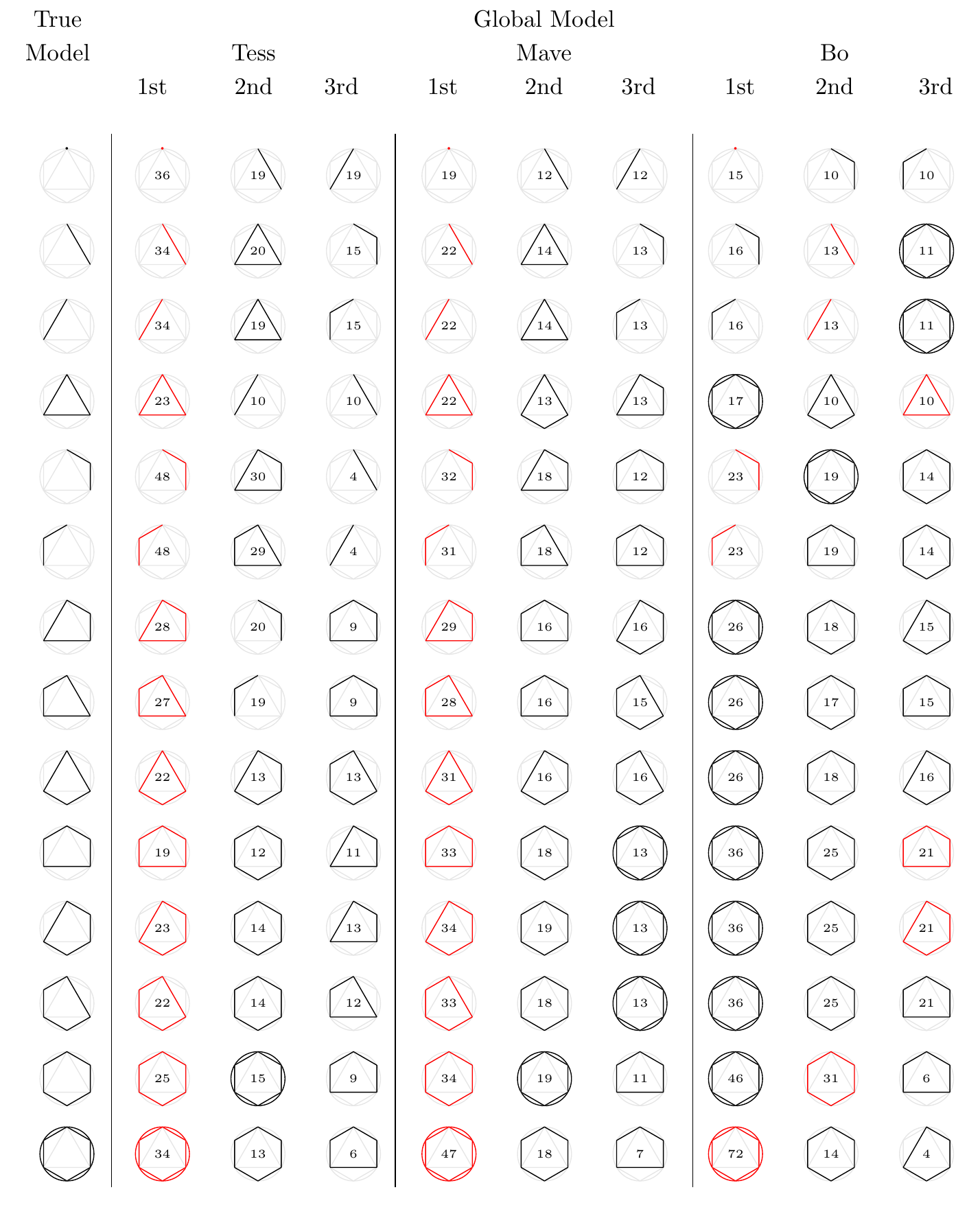}
\captionsetup{labelformat=empty}
\caption{{\bf S13 Figure. The three most visited models for scientist populations with one dominant type and the proportion of time spent at each true model, when AIC is the model comparison statistic and noise equals the signal in a system with no replication.} For $\sigma^2:\E(y|\mu_x)=(1:1)$, proportion of time spent by a model as the global given a true model, assessed by AIC. Three most visited models are shown. Numbers show time spent at each model in percent points. True models are in red. \textit{Tess-} and \textit{Mave-}dominant populations perform more poorly than they do under low error, however, they still spend more time at the true model than any other models. Surprisingly, \textit{Bo-}dominant population captures the true model more often now than under low noise although it still performs relatively poorly as compared to other homogeneous populations.}
\label{S13_Figure}
\end{figure}

These results from the system with no replication show that while the true model is sticky and reached quickly under low error in a well connected system, the scientific population still spends considerable time in false models over the long run. Moreover, proportion of research strategies in scientific populations, true model complexity, and model comparison statistic have an effect on all of these properties. Overall, \textit{Bo-}dominant population performs poorly in most scenarios whereas \textit{Tess-} and \textit{Mave-}dominant populations excel in different scenarios. Epistemically diverse population minimizes the risk of worst outcomes. These patterns that we described change substantially as the ratio of error variance to model expectation in the system increases (\hyperref[S13_Figure]{S13 Figure}-\hyperref[S22_Figure]{S22 Figure}). We now discuss the implications and limitations of results presented so far for the scientific practice.

\textbf{Implications and limitations.} When the truth exists and is accessible, we find that scientific process indeed discovers and sticks to it in most situations. The exceptions to this result come from 1) research strategies that search the model space in a biased manner and fail to test the true model against alternatives often enough, and 2) large error in the data generating process. In practice, (1) might be realized when there is no overarching theoretical framework guiding the search of model space but instead folk theories or intuitions are used to reduce the possibilities~\cite{Muthukrishna2019}. Further, (2) is a real challenge, especially in disciplines where data do not carry a strong signal (e.g., low estimated effect sizes in psychology literature) or whose methods are not sufficiently fine-tuned to detect the signal (e.g., high measurement error). While these are implications that might hold qualitatively for real-life scientific practice, we caution the readers to not over-generalize specific parameters such as the stickiness of a true model and the proportion of time spent at the true model. These quantities depend on the parameters of our system, such as the number of models in the universe, and the linear models framework.

\subsection*{Results in a system with replication}
\label{sec:ABM} In addition to the properties analyzed in the previous section, in the system with replication we can also analyze the rate of reproducibility since we introduce a replicator in the system. One of our goals is to understand the relationship between reproducibility and other desirable properties of scientific discovery. Informed by the findings reported in the previous section, we run the ABM under three true models of varying complexity, three levels of error variance to model expectation ratio, five scientific populations, and two model comparison statistics (\hyperref[S9_Table]{S9 Table}). Moreover, ABM allows us to implement hard research strategies where scientists propose only models complying with their strategies and all models incompatible with a given research strategy have zero probability of being proposed by the scientists pursuing that strategy (\nameref{S5_Appendix}). Thus, connectedness among models is restricted in this framework for all scientists with the exception of \textit{Mave} whose research strategy allows her to propose any model at any given time, and who thereby maintains a soft research strategy. When the transition matrix is highly connected, the discovery of truth is fast, as shown in the previous section. In the current section, we explore how the speed of discovery changes under restricted connectedness for different scientist populations.
\subsubsection*{Reproducible results do not imply convergence to scientific truth} We first explored the relationship between the rate of reproducibility and other desirable properties of scientific discovery, and found that this relationship must be interpreted with caution. In our framework, we defined the rate of reproducibility as the probability of the global model staying the same after a replication experiment. We show that the rate of reproducibility has no \textit{causal} effect on other desirable properties of scientific discovery including: the probability that a model is selected as the global model in the long run, the mean first time to hit a model, and stickiness of a model (see~\nameref{S23_Appendix} for mathematical proof). Thus, although multiple confirmations of a result in a scientific inquiry lend credibility to that result, withstanding the test of multiple confirmations is not sufficient for convergence to scientific truth.

\begin{figure}[!ht]
\includegraphics[width=1\textwidth]{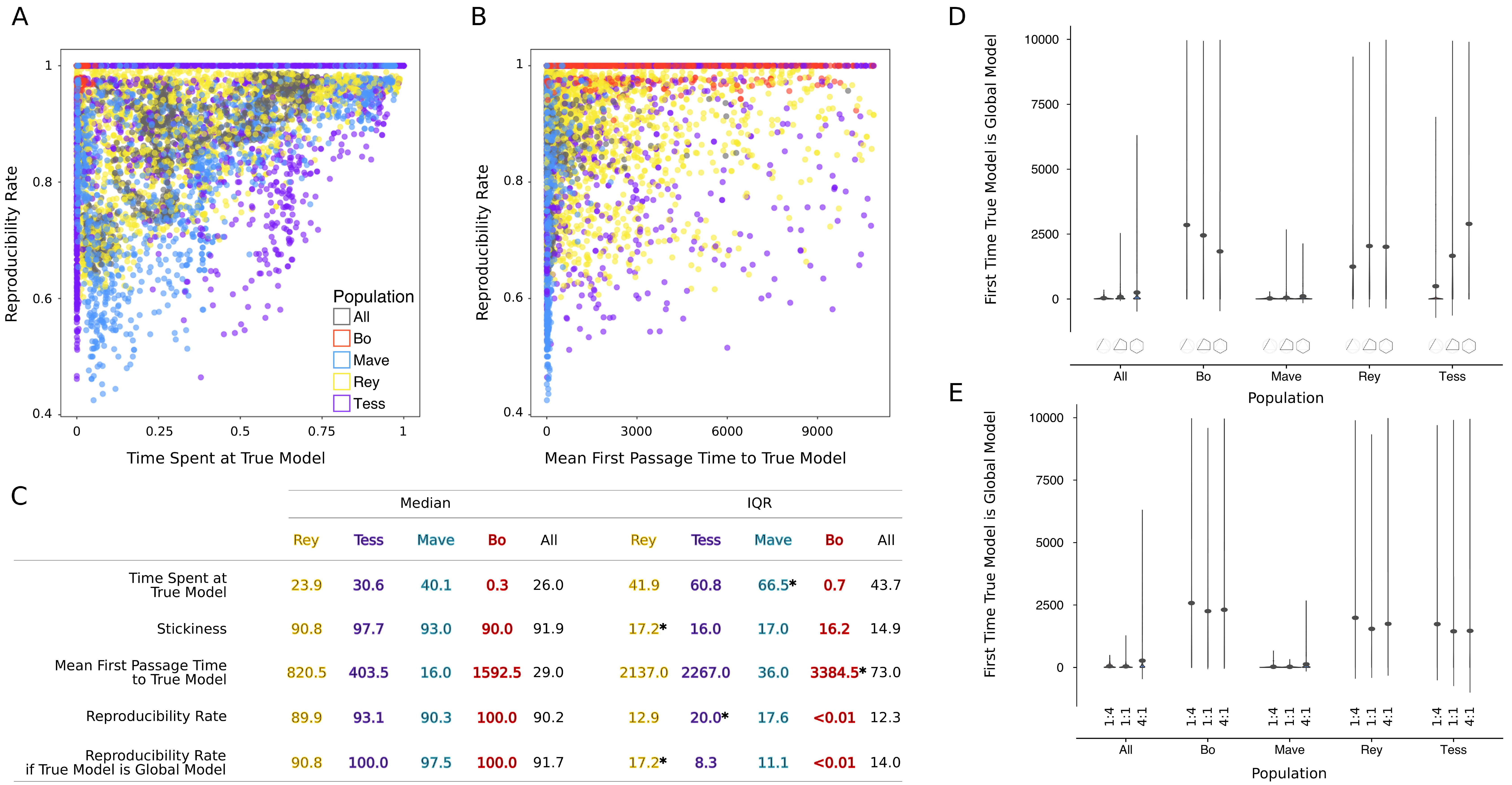}
\caption{For \textit{Rey}-, \textit{Tess}-, \textit{Mave}-, \textit{Bo}-dominant, and epistemically diverse populations: (A) The rate of reproducibility against time spent at true model. (B) The rate of reproducibility against mean first passage time to true model. (C) Summary statistics with highest IQRs indicated by $*$. Mean first passage time to true model in number of time steps; all else in percent points. Violin plots for the mean first passage time to the true model per population type versus (D) complexity of true model and (E) error variance to model expectation ratio. Dots mark the means.}
\label{fig:ABM1}
\end{figure}

On the other hand, desirable properties of scientific discovery and the rate of reproducibility might be correlated. Whether there is any correlation depends on the research strategies and their frequency in the population (see~\nameref{S23_Appendix} for mathematical explanation). We present scatter plots (Fig.~\ref{fig:ABM1}A and Fig.~\ref{fig:ABM1}B) as evidence for the complexity of these correlations across scientist populations. From these scatter plots and Fig.~\ref{fig:ABM1}C, we see, for example, that \textit{Bo-}dominant populations reach perfect rate of reproducibility while spending little time at the true model (as assessed by the very low Spearman rank-order correlation coefficient, $r_{SR}=-0.06$), which confirms that high rate of reproducibility does not imply true results.

Across all simulations, as the rate of reproducibility increases, scientist populations do not necessarily spend more time on the true model, as indicated by a lack of correlation between rate of reproducibility and time spent at the true model ($r_{SR}=-0.02$, Fig.~\ref{fig:ABM1}A). Further, as the rate of reproducibility increases, the discovery of truth slows down rather than speeding up as shown by a positive but small correlation ($r_{SR}=0.26$, Fig.~\ref{fig:ABM1}B). Crucially, both of these correlations are driven by the research strategy dominant in the population and should only be taken as evidence for the complexity of these relationships between rate of reproducibility and other desirable properties of scientific discovery (see Table~\ref{S24_Table} in~\nameref{S24_Appendix} for all correlation coefficients per scientist population). For example, \textit{Bo}-dominant population reaches almost perfect reproducibility (Fig.~\ref{fig:ABM1}A-C, red) while taking a long time to hit the true model and spending short time at it. On the other hand, \textit{Mave}-dominant population hits the true model quickly and spends long time there (Fig.~\ref{fig:ABM1}A-C, blue), but it has a much lower rate of reproducibility than \textit{Bo}-dominant population.

\textbf{Implications and limitations.} This counter-intuitive result on reproducibility is due to violating assumptions of statistical methods. Statistical theory guarantees to find the true data generating mechanism as the best fit for the observed data, if a reliable method operates in the absence of external factors. The research strategies implemented in our ABM include critical external factors that determine how the model space is searched. For \textit{Bo}-dominant populations, we get high levels of reproducibility and low level of actual discovery for the same reason: Models proposed by \textit{Bo}s consistently result in fitting overly-complex models to data but lead to reproducible inferences since the comparisons are often between models that are 1) not true and 2) far from each other in the sense of statistical distance. As a result, even though the true model is not proposed, and hence not selected as the global model, the method consistently favors the same untrue model when a specific comparison is repeated with independent data.

\textit{Mave}-dominant populations search the full model space and consequently discover the true model quickly. Their rate of reproducibility is lower relative to \textit{Bo}-dominant populations. This is because randomly proposed models are typically not true, but also, \textit{Mave}s do not have a biased strategy of proposing models that are far away from the global model. Hence, \textit{Mave}'s comparisons do not always favor a specific model especially when models close to each other are tested.

The effects discussed in this section are marginal main effects of scientist populations over other factors that we vary in our ABM, including levels of noise-to-signal ratio. Due to this marginalization, the mean noise-to-signal ratio that affects the results is higher in our ABM than the system without replication. Nonetheless, our sanity checks still capture the salient properties of scientific discovery well. For example, the basic expectation that a successful scientific endeavor will move us closer to truth is captured. Fig.~\ref{fig:ABM1}C confirms that most scientific populations (with the exception of \textit{Bo}-dominant) spend considerable time at the true model. In all populations, the true model is sticky. The rate of reproducibility under the true model is higher than the overall rate of reproducibility across all models that become the global model. The scientific community ultimately discovers the true model with varying speed, depending on dominant strategies represented in the population. We also find that the rate of reproducibility is highly positively correlated with stickiness of the true model in most scientist populations except \textit{Bo}-dominant (Table~\ref{S24_Table} in~\nameref{S24_Appendix}). The rate of reproducibility is also positively correlated with time spent at true model for most populations, although these correlations are expectedly lower because they are unconditional on the first time to discovery. Speed of discovery has low correlation with the rate of reproducibility. This makes intuitive sense because speed of discovery is largely determined by how the model space is searched (i.e., an external factor) whereas the rate of reproducibility is collectively determined by all variables in the system.

Against this backdrop, we speculate how these counter-intuitive findings might extend to the practice of science beyond our theoretical framework. What we observe appears akin to the tension noted by Shiffrin, B\"{o}rner, and Stigler~\cite{Shiffrin2018} regarding the risk of obstructing scientific exploration by imposing restrictions on how science should be conducted. Indeed, we show that exploratory strategies (represented by \textit{Mave}s in our system) are needed to speed up scientific discovery. But then we also need scientists testing theory and running replication studies (e.g, \textit{Tess} and \textit{Rey}s) to establish which discoveries are \textit{true}. If we restrict exploration to allow only research strategies that search the model space in an extremely biased manner (e.g, \textit{Bo}s), we may lock ourselves in a vicious circle of never making a discovery. The reason is that we may be able to obtain high rate of reproducibility as an artifact of this research strategy. Our cases may be extreme and in reality we might expect diverse scientist populations to emerge naturally. However, past research suggests that if incentive structures reward \textit{Bo}-like strategies due to high rates of reproducibility they report, these strategies may be widely selected for in scientific populations~\cite{Smaldino2016} thereby resulting in canonization of false results~\cite{Nissen2016}.

\subsubsection*{Innovation speeds up scientific discovery} \textit{Mave}-dominant population is the fastest to hit the true model (Fig.~\ref{fig:ABM1}C \hyperref[S25_Figure]{S25 Figure} A) regardless of the true model complexity and the error variance to model expectation ratio (Fig.~\ref{fig:ABM1}D and Fig.~\ref{fig:ABM1}E). Further, for epistemically diverse population in which all scientist types are equally represented, the proportion of mavericks is sufficient to garner this desirable result. The reason is that \textit{Mave} provides connectedness in transitioning from model to model via her soft research strategy even when all other scientists represented in the scientific population pursue hard research strategies. All other homogeneous populations take a long time to reach the truth due to pursuing hard research strategies. For example, the estimate for mean first passage time to the true model for \textit{Bo-}dominant population is $1592.5$ steps (Fig.~\ref{fig:ABM1}C). We also ran the ABM with soft research strategies and include the results regarding speed of discovery in Table~\ref{S26_Table} (\nameref{S26_Appendix}) as further confirmation that connectedness among models leads to faster discovery in \textit{Tess}- and \textit{Bo}-dominant populations, besides \textit{Mave}-dominant and epistemically diverse populations.

\textbf{Implications and limitations.} The idea that innovative research plays a significant role in scientific discovery is intuitive and hardly new~\cite{Weisberg2009, Muldoon2013, Alexander2015}. Our results qualify this idea in a particular way: Innovation leads to fast discovery, which is a property determined by the stochastic process governing the connectedness of models. We should note that the memorylessness property of our system might have exaggerated the role of \textit{Mave}s in making a quick discovery. If all scientists carry a tally of past results and adjust their strategies accordingly, it is possible that the model space could be explored more efficiently by scientist types other than \textit{Mave}. What is needed in essence is not \textit{Mave}s necessarily but a way to guarantee high connectedness among models in the search space and an efficient search algorithm. Arguably the role of innovative, exploratory research is more critical early on in the research cycle and once we are in the vicinity of truth, limited scientific resources might be better spent elsewhere (e.g., confirmatory research or pursuit of other research questions).

\subsubsection*{Epistemic diversity optimizes the process of scientific discovery}
\begin{figure}[H]
\centering
\includegraphics[width=1\textwidth, angle=90]{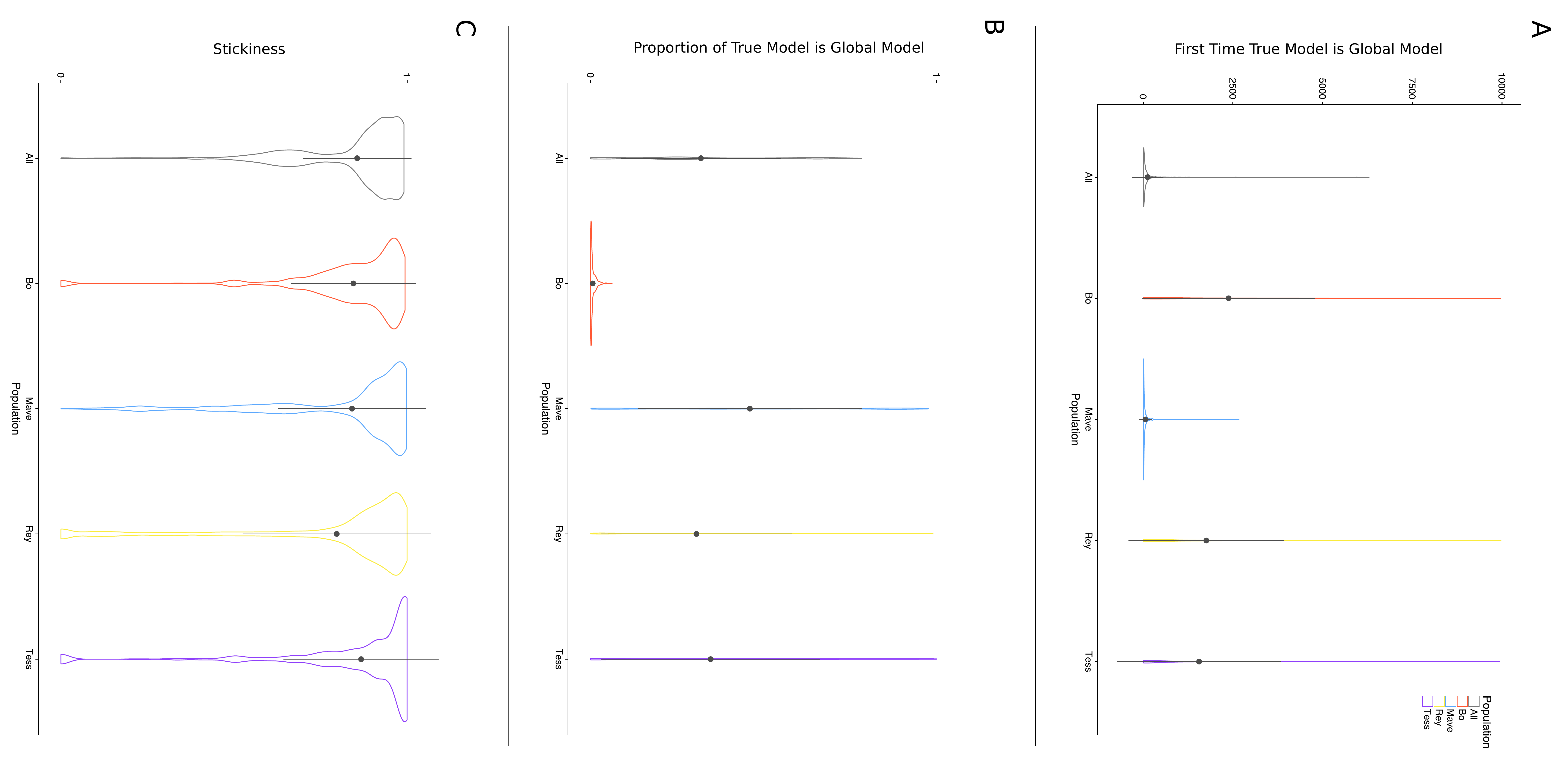}
\captionsetup{labelformat=empty}
\caption{{\bf S25 Figure. A comparison of all five scientist populations in ABM on three properties of scientific discovery in a system with replication.} Violin plots showing marginal effects of scientist populations on first passage time to true model (A), proportion of times true model is global model (B), and true model stickiness (C). While different homogeneous populations appear to perform better/worse on different properties, epistemically diverse population (indicated by All) appears to have lowest variability across outcomes.}
\label{S25_Figure}
\end{figure}

We looked at which scientific population optimizes across all desirable properties of scientific discovery. Figure~\ref{fig:ABM1}C summarizes the sample median and interquartile range for the time spent at, the stickiness of, and the mean first passage time to the true model, as well as the rate of reproducibility (also see~\hyperref[S25_Figure]{S25 Figure}). These statistics show the advantage of an epistemically diverse population of scientists on the efficiency of scientific discovery. Homogeneous populations with one dominant research strategy tend to perform poorly in at least one of these desirable properties. For example, \textit{Rey}-dominant population has low median rate of reproducibility. \textit{Mave}-dominant population has low median rate of reproducibility and high variability in time spent at the true model. \textit{Tess}-dominant population has high variability in mean first passage time to the true model and the rate of reproducibility. \textit{Bo}-dominant population has low median time spent at the true model, low median stickiness, and high variability in mean first passage time to the true model. In contrast to all these examples, epistemically diverse population \textit{always} performs better than the worst homogeneous population with respect to \textit{all} properties and further, it has low variability. Thus, epistemic diversity serves as a buffer against weaknesses of each research strategy, consistent with results from the system with no replication. We conclude that among the scientist populations we investigate, epistemic diversity optimizes the properties of scientific discovery that we specified as desirable.

\textbf{Implications and limitations.} We believe that the importance of epistemic diversity is intuitive, yet, it cannot be emphasized enough. Our definition of epistemic diversity is limited to the representation of the four research strategies that we included in our system. In reality, there are numerous philosophical (e.g., logical positivist, post-modernist), research methodological (e.g., empirical experimentation, computer simulations, ethnography), and statistical (e.g., frequentist, likelihoodist, Bayesian) approaches to conducting science and our model is agnostic as to what kind or what degree of epistemic diversity would optimize scientific discovery. We merely find that the role of epistemic diversity in scientific population is akin to diversifying an investment portfolio to reduce risk while trying to optimize returns.

\subsubsection*{Methodological choices affect time spent at scientific truth} 
\begin{figure}[H]
\centering
\includegraphics[width=1\textwidth]{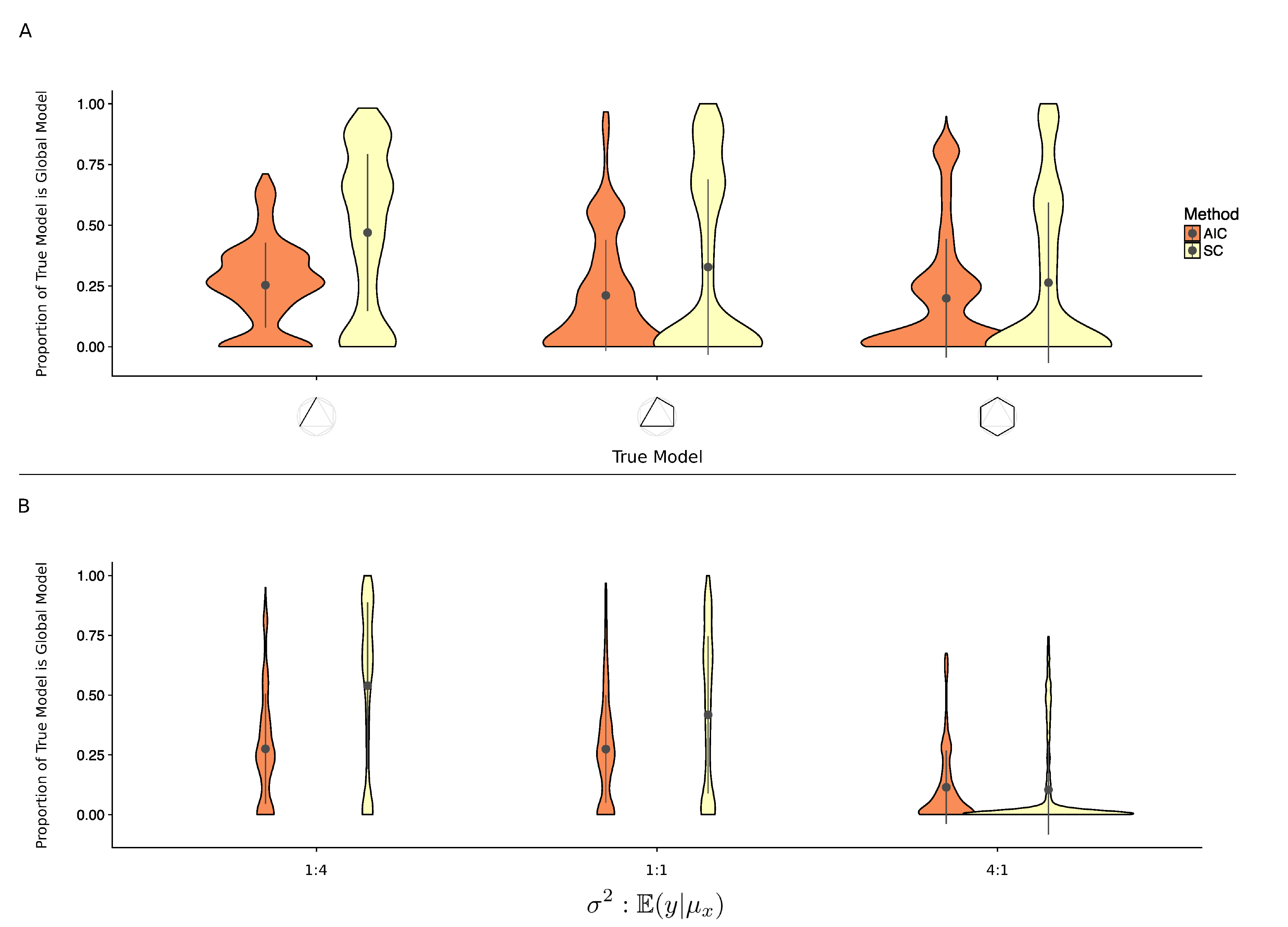}
\captionsetup{labelformat=empty}
\caption{{\bf S27 Figure. Interaction of model comparison statistics with true model complexity (A) and with error variance to model expectation ratio (B) on time spent at the true model in a system with replication.} (A) Violin plots for proportion of times true model is global model per model comparison statistic and complexity of true model. (B) Violin plots for proportion of times the true model is global model per model comparison statistic and the ratio of error variance to model expectation. Scientist populations spend more time at the true model under SC than AIC when the model is simple or when error variance to model expectation ratio is low ($1:4$). Time spent at true model decreases and difference between AIC and SC disappears as model complexity or error variance to model expectation ratio increases.}
\label{S27_Figure}
\end{figure}

The choice of method may appear to be perfunctory if multiple methods perform well. However, violating the assumptions of a method affects the results of an analysis performed with that method. The effects of the model comparison statistic in our system, where a comparison of misspecified models is routinely performed, is not trivial~\cite{Lv2013}. When true model complexity is low, using SC for model comparison increases the time spent at the true model compared to AIC (\hyperref[S27_Figure]{S27 Figure}A). As model complexity increases, however, this difference disappears and further, AIC has lower variability. When the ratio of error variance to model expectation is low, SC leads to a longer time spent at the true model. As the ratio of error variance to model expectation increases, AIC and SC spend comparable amount of time at the true model, but AIC has smaller variability (\hyperref[S27_Figure]{S27 Figure}B). Averaged over all other parameters, SC spends longer time at the true model than AIC (\textit{medians} = $27.05\%$ and $19.83\%$, respectively), but with greater variability (\textit{IQR} = $66.03\%$ and $33.80\%$, respectively).

\textbf{Implications and limitations.} The finding that methodological tools might affect scientific progress is factually known~\cite{Box1976, Nelder1986} and being studied extensively by statisticians and meta-scientists alike. Model comparison methods such as AIC and SC as well as all other statistical inference methods work best when their assumptions are met and might lead to invalid inferences under assumption violations. An unsurprising implication of our findings is that statistical theory should inform statistical practice even in the absence of well-known procedural violations such as p-hacking.
\section*{Conclusion}
\label{sec:discussion} We studied the process of scientific discovery and reproducibility in a meta-scientific framework using a model-centric approach.
We have chosen a model-centric approach because 1) it translates to scientific models directly, 2) it is a generic mode of inference encompassing hypothesis testing, and 3) model selection methods bypass difficulties associated with classical hypothesis testing.

Our scientists engage in straightforward research strategies and do not commit experimenter bias, learn from their own or others' experiences, engage in hypothesis testing, or commit measurement errors. Further, they are not prone to QRPs or structural incentives. We also assume that there exists a true model that our scientist population attempts to discover and that this true model is within the search space readily available to the scientist population. These factors that we have abstracted away are potential avenues for future research, particularly for complex social dynamics, but our goal here was to explore how the process of scientific discovery works in an idealized framework. We did, however, provide sanity checks to make sure that our system behaves in meaningful ways with respect to what we would expect from a well-functioning scientific process.
\par Our study shows that even in this idealized framework, the link between reproducibility and the convergence to a scientific truth is not straightforward. A dominant research strategy producing highly reproducible results might select untrue models and steer the scientific community away from the truth. Reproducible false results may also arise due to bias in methods and instruments used, as discussed by Baumgaertner and colleagues~\cite{Baumgaertner2018}. While both reproducibility and convergence to a scientific truth are presumably desirable properties of scientific discovery, they are not equivalent concepts.
In our system inequivalence of these concepts is explained by a combination of research strategies, statistical methods, noise-to-signal ratio, and the complexity of truth. This finding further indicates that issues regarding reproducibility or validity of scientific results should not be reduced down to QRPs or structural incentives. Considering such methodological and institutional factors, however, would add additional layers of complication, moving us even further away from the guarantees provided by statistical theory.
\par Not all our results are as counter-intuitive however. On a positive note, we find that the process of scientific discovery is rendered efficient if the transitions between models in the model space are easy. Scientist populations that expedite transitions via  promoting innovative research or pursuing flexible strategies will discover the truth quickly. In real life, we surmise that the model space might be much larger and the true model--if it exists--might not necessarily be easily accessible in the search space. Therefore, an outstanding challenge for science appears to be to attain a scientific population that can realize optimum connectedness in the model space to expedite the discovery of truth.
\par Recently, Shiffrin, B\"{o}rner, and Stigler~\cite{Shiffrin2018} have warned against ``one size fits all'' type of approaches in science and scientific reforms, advising a nuanced approach instead (p.2638). Complementary to their perspective, our results also advise against homogeneity in scientific practice. We find that a diversity of strategies in the scientific population optimizes across desirable properties of scientific discovery---a finding consistent with the cognitive division of labor literature~\cite{Zollman2009}. If populations are largely homogeneous, with one research strategy dominant over others, then the scientific population tends to perform poorly on at least one of the desirable properties which might mean forsaking reproducibility or delaying discovery.
\par We find that the choice of statistics relative to true model complexity has non-trivial effects on our results. This is corroborated by recent statistical theory~\cite{Lv2013}. The difficulty is that the complexity of the true model is often unknown to scientists who make not only their statistical inference but also their methodological choices under uncertainty. We believe that model complexity may have differential effects on the desirable properties of scientific discovery depending on the choice of statistic.
\par Our model, as any other model, is an abstraction of reality and we believe that we have captured salient features of the scientific process of interest to our research questions. Main limitations of our framework are the lack of capacity to learn and memorylessness of scientists. The replicator only provides meta-level information about the scientific process and does not contribute directly to the accumulation of scientific knowledge. Incorporating past reproducibility of specific results in decision making strategies might allow the replicator to make substantial contributions to scientific discovery. A realistic implementation of this aspect requires our virtual scientists to adopt machine learning algorithms that can heuristically teach them to become intelligent agents.

Our research also raises questions with regard to reproducibility of scientific results. If reproducibility can be uncorrelated with other possibly desirable properties of scientific discovery, optimizing the scientific process for reproducibility might present trade-offs against other desirable properties. How should scientists resolve such trade-offs? What outcomes should scientists aim for to facilitate an efficient and proficient scientific process? We leave such considerations for future work.
\section*{Supporting information}




\paragraph*{S1 Appendix.}
\label{S1_Appendix}
{\bf A stochastic process of scientific discovery.}
Consider an infinite population of scientists conducting a sequence of idealized experiments $\xi^{(t)}:=(M_P^{(t)},\theta, D^{(t)},S,K^{(t)})$, indexed by time $t=1,2,\cdots$ where $M_P^{(t)}$ belongs to a set of probability structures $\M=\{M_1,M_2,\cdots,M_L\}$ known to all scientists. Further, assume that there are $A$ distinct scientist types in the population, each with a well-defined research strategy $R\in \R=\{R_o,R_1,\cdots,R_A\}$ of proposing a model in their experiment. These strategies depend on the type of scientist and a global model $M_G^{(t)} \in \M, K^{(t)},$ which represents the consensus of the scientist population at time $t$. The population of scientists aims to find the true model $M_T \in \M$. A scientist selected to conduct an experiment at time $t$, uses her background knowledge $K^{(t)}$ to propose a new candidate model $M_P^{(t)}$. Specifically, we define $K^{(t)}$ as a probability distribution $\Prob(M_P|R^{(t)},M_G^{(t)})$, where $\{M_P,M_G^{(t)}\}\in \M^2$, and $R^{(t)} \in \R$.

The initial conditions of our stochastic process include the true model $M_T$, true parameter values $\theta_T$ of $M_T$, an initial global model $M_G^{(0)}$, a method for model selection $S$, and the sample size of the data $n$. At each time step, an idealized experiment $\xi^{(t)}$ is performed and new data $D^{(t)}$ of size $n$ is generated independent of everything else from distribution $M_T(\theta_T)$. Each experiment is performed by a scientist randomly selected from $A$ types in the population using the categorical distribution with probabilities $(p_{1},p_{2},\cdots,p_{A})$. The selected scientist proposes a model $M_P^{(t)}$ with probability $\Prob(M_P|R^{(t)},M_G^{(t)})$ conditional on a research strategy fully specified by her type and the current global model. Given the data $D^{(t)}$, model scores under the proposed model and the current global model are calculated as $S(M_P^{(t)})$ and $S(M_G^{(t)})$, respectively. The model with favorable score (i.e., smaller for both AIC and SC) is set as the new global model $M_G^{(t+1)}$. This mechanism represents how scientific consensus is updated in light of new evidence.

A defining property of our stochastic process with no replication is that $K^{(t)}$ depends only on quantities at time $t$. If $R_a\in \mathcal{R}$ depends only on $M_G^{(t)}$ for all $a$, the transition from $M_G^{(t)}$ to $M_G^{(t+1)}$ admits the Markov property and the stochastic process representing the scientific process is a Markov chain with transition probabilities given by
\begin{equation}\label{eq:transitionprobs}
\Prob(M_G^{(t+1)}=M_\ell|M_G^{(t)}=M_i)=
\sum_{a=1}^{A}\Prob(S(M_\ell)<S(M_i))\Prob(M_\ell|R_a,M_i)\Prob(R_a).
\end{equation}
On the right hand side of Eq.~\eqref{eq:transitionprobs}, the last term is the probability of selecting a scientist with research strategy $R_a$ independent of all else, the middle term is the probability of proposing the model $M_\ell$ given the current global model $M_i$ and the scientist type $a$ with research strategy $R_a$ selected. The probability $\Prob(S(M_\ell)<S(M_i))$ depends on $M_T$ via $D^{(t)}$ generated and it is obtained by
%
$
\int_{\Theta}\int_{\mathcal{D}}\Prob(S(M_\ell)<S(M_i)|D)\Prob(D|\theta)\Prob(\theta)dDd\theta,
$
%
where $\Prob(\theta)$ is the probability of parameter, $\Prob(D|\theta)$ is the likelihood of the data, and $\Prob(S(M_\ell)<S(M_i)|D)$ is the probability that the proposed model $M_\ell$ has a more favorable score than $M_i$ conditional on data. By convention we set $\Prob(S(M_\ell)<S(M)_i)=1$ when $\ell=i,$ and $\Prob(M_\ell|R_a,M_i)>0$
for all $a,i,\ell$ so that transition probabilities are nonzero for all models and scientist types. This second condition guarantees that our Markov chain is ergodic, which implies that it has a unique stationary distribution---its limiting distribution for visiting a model.

When there are no replication experiments in the system, $K^{(t)}$ is defined as $\Prob(M_P|R^{(t)},M_G^{(t)})$ which states that conditional on $R^{(t)}$ and $M_G^{(t)}$, the probability of proposing a model is independent of the past time steps. Let $R_o\in \mathcal{R}$ be the replicator strategy. Given the proposed and global models at time $t-1$, the replicator strategy at time $t$, $R_o^{(t)}$, is to perform an experiment at time $t$, using the exact same proposed and global models as those at time $t-1$, but with new data $D^{(t)}$ generated under $M_T(\theta_T)$. Since $R_o\in R$ depends on $M_G^{(t-1)}$, the transition from $M_G^{(t)}$ to $M_G^{(t+1)}$ does not admit the Markov property anymore and the stochastic process representing the scientific process is a higher order Markov chain. The transition probabilities of the Markov chain at time $t$ can be expressed by conditioning on whether a scientist chosen at a given time is a replicator:
\begin{eqnarray}\label{eq:1}
\nonumber
& &\Prob(R^{(t)}\neq R_o)\Prob(M_G^{(t+1)}|M_G^{(t)})+\\\nonumber
& &\Prob(R^{(t)}= R_o)[\Prob(R^{(t-1)}\neq R_o)\Prob(M_G^{(t+1)}|M_G^{(t)},
M_G^{(t-1)})+\cdots+\\
& &\Prob(R^{(1)}= R_o)[\Prob(R^{(0)}\neq R_o)\Prob(M_G^{(t+1)}|M_G^{(t)}, M_G^{(t-1)},\cdots,M_G^{(0)})]\cdots].
\end{eqnarray}
In Eq.~\eqref{eq:1}, the first term in the sum is the joint probability of choosing a scientist who is not a replicator at time $t$ and the transition probability from global model at time $t$ to global model at time $t+1$. Since the scientists are chosen independently of all else, the joint probability is written as the product of choosing a scientist who is not a replicator at time $t$, given by $\Prob(R^{(t)}\neq R_o)$, and the probability of transition to the global model at time $t+1$ is given by Eq.~\eqref{eq:transitionprobs}. The second term in the sum is the joint probability of choosing a scientist who is a replicator at time $t$ and the transition probabilities to a model. We write the second term as the product of $\Prob(R^{(t)}=R_o)$ and the transition probabilities when a replicator is chosen. If the scientist at time $t$ is a replicator, she replicates the experiment at time step $t-1$, which might be a replication experiment itself. Therefore, the transition probabilities to a model within the first brackets is a sum of two probabilities. The first term is the joint probability of choosing a scientist who is not a replicator and the transition probability in that case, and the second term is the probability of choosing a replicator given by $\Prob(R^{(t-1)}=R_o)$ at time step $t-1$, and the transition probability in that case. This is a recursive equation, in the sense that the transition probabilities at time $t$ depend on the transition probabilities at time $t-1$. An implication is that the transition probabilities at time $t$ are path dependent. Therefore, when a replicator scientist is included in the population, we have a higher order Markov chain, whose long term dynamics are feasible to obtain with a forward simulation method.

For the process with replicator, we lift the assumption $\Prob(M_\ell|R_a,M_i)>0$ for all $a,i,\ell$ that we imposed in the process without a replicator. This assumption increases the connectivity of the transition probability matrix, which makes calculations in the long-term behavior of the Markov chain straightforward. Due to our new process not admitting the Markov property, these calculations are irrelevant in the analysis of the process with a replicator. Therefore, we drop the assumption of transitioning from a model to any other model to be nonzero. Removing this assumption allows us to define scientist types that visit only the subset of all models consistent with a specific research strategy. This property of the process renders the effects of each research strategy on the process outcomes well-pronounced.
\paragraph*{S2 Appendix.}
 \label{S2_Appendix}
{\bf Description of example system of linear models.}
We define $\M$, the family of linear models as
$$\{\mathbf{y}=\mathbf{X}\bm{\beta}+\bm{\epsilon}|\beta_{i\cdots jv}\neq 0 \Rightarrow (\beta_{i}\neq0,\cdots,\beta_{v}\neq0, \cdots, \beta_{i\cdots j}\neq0)\}.$$
Here, $\mathbf{y}$ is $n \times 1$ vector of response variables, $\mathbf{X}$ is $n \times p$ matrix of predictor variables, $\bm{\beta}$ is $p \times 1$ vector of model parameters, and $\bm{\epsilon}$ is $n \times 1$ vector of random errors satisfying the Gauss-Markov conditions $\E(\epsilon_i)=0,{\rm Var}(\epsilon_i)=\sigma^2$, and ${\rm Cov}(\epsilon_i,\epsilon_j)=0$ for all $i,j$. $\M$ contains $L$ linear models with up to $k$ factors subject to the constraint $\beta_{i\cdots jv}\neq 0 \Rightarrow (\beta_{i}\neq0,\; \cdots, \beta_{i\cdots j}\neq0)$ which guarantees that if a model contains a $v$--way $(v\leq k)$ interaction term of $v$ factors in the model, then all $(v-1)$--way, $(v-2)$--way, $\cdots$, $2$--way interactions of those factors and their main effects are included in the model. We include the predictor $x_1$ and the response variable $y$ in all models reflecting our assumption that all scientists in the community focus on a research question that involves at least one common factor of interest and a common response variable.
%
%
Let $M_i \in \M$ be a model with $p_i$ parameters of which $v_{\ell_i}$ is the highest order interaction term with order $\ell_i$ denoting the order, $\#{v_{\ell_i}}$ is the cardinality of $v_{\ell_i}$. Let $M_i \succ M_j$ denote that $M_i$ is more complex than $M_j$.  We define the \textit{model complexity} as a partial ordering obeying three conditions:
\begin{enumerate}
\item If $p_i>p_j$ then $M_i\succ M_j$.
\item If $p_i=p_j$ and $\ell_i>\ell_j$, then $M_i\succ M_j$.
\item If $p_i=p_j,\;\ell_i=\ell_j$, and $\#{v_{\ell_i}}>\#{v_{\ell_j}}$, then $M_i\succ M_j$.
\end{enumerate}
Otherwise, we say that complexities of $M_i$ and $M_j$ are indistinguishable.

\paragraph* {S3 Algorithm.}
 \label{S3_Algorithm}
{\bf Agent-based model algorithm.}
We first present the general algorithm in the agent-based model and then discuss specific values used in the article.
 \begin{algorithm}[!ht]
   \caption{Algorithm for stochastic process of scientific discovery with replicator.}
   \begin{algorithmic}[1]
     \STATE Input: $\M, \Theta, S, \R, \Prob(R_a), \Prob(M|R_a,M_G^{(t)}), M_G^{(0)}, M_T, \theta_T, n, t_{max}$
     \STATE Set $t=0$
     \WHILE{$t < t_{max}$}
       \STATE Simulate $R_a\sim {\rm Categorical}(p_{1},p_{2},\cdots,p_{A})$
       \STATE Simulate $M_P^{(t)}\sim \Prob(M|R_a,M_G^{(t)})$
       \STATE Simulate $D^{(t)}_i\sim M_{T}(\theta_T)$, for $i=1,2,\cdots,n$ independently of each other
       \STATE Calculate
         \begin{align*}
           S(M_P^{(t)})-S(M_G^{(t)})&=C+\sum_{i=1}^{n}\log \Prob(D^{(t)}_i|\hat{\theta},M_P^{(t)})-\sum_{i=1}^{n}\log\Prob(D^{(t)}_i|\hat{\theta},M_G^{(t)}),
         \end{align*}
         where, $C=2p\log(n)$ if SC, or $C=2p$ if AIC, and $\hat{\theta}$ is the maximum likelihood estimate of $\theta$
       \IF{$S(M_P^{(t)})<S(M_G^{(t)})$,}
         \STATE Set $M_G^{(t+1)}=M_P^{(t)}$,
       \ELSE
         \STATE Set $M_G^{(t+1)}=M_G^{(t)}$
       \ENDIF
       \STATE Set $t = t + 1$
     \ENDWHILE
   \end{algorithmic}
 \end{algorithm}

We choose $M_G^{(0)}$ randomly with equal probability from models in $\M$. $\Theta$ determines $\theta_{min}, \theta_{max}$ and $\theta_T$ is chosen uniformly randomly on this interval. The parameters of the categorical distribution used in step 4 is determined by the proportion of scientists in the population.
\newpage
\paragraph*{S4 Appendix.}
 \label{S4_Appendix}
 {\bf Properties of scientific discovery.}

{\bf How quickly does scientific community discover the true model?} When there are no replicators in the model, $\mathbf{P}=\{\Prob(M_\ell|M_i)\}$ is the transition probability matrix given by Eq.~\eqref{eq:transitionprobs}. We assess the speed with which scientific community discovers the true model by the mean first passage (or hit) time to the true model. This is the first time to reach $M_T$, given that we start the system from $M_i$. The mean first passage time to the true model which we denote by $\tau_{i,T}$, is the expected value of this first passage time to select the true model as the global model for the first time in the process, given that the process starts from a known model $M_i$. By theory of Markov chains we have
\begin{equation}
\tau_{i,T}=1+\sum_{\substack{\ell=1\\M_\ell\neq M_T}}^L\Prob(M_\ell|M_i)\tau_{\ell,T},\; i=1,2,\cdots, L.\nonumber
\end{equation}
Given the transition probabilities $\Prob(M_\ell|M_i)$, the solution to this system of $L$ linear equations is readily obtained by a method to solve linear equations.

When there are replicators in the model, we assess the speed to discover the true model by the mean number of steps for the system to update the global model to the true model for the first time. The first the time true model is global model is the mean first passage time to the true model described in the process of scientific discovery with no replication, but it is unconditional on the starting model.

{\bf How ``sticky'' is the true model as global model?} We define stickiness of a model as the mean---over proposed models---probability of staying in model $M_i$ conditional on the current global model $M_i$. When there are no replicators in the model, by theory of Markov chains, it is given by
\begin{equation}\label{eq:sticky}
\Prob(M_G^{(t+1)}=M_i|M_G^{(t)}=M_i)=\sum_{a=1}^{A}\sum_{\ell=1}^{L}
\Prob(S(M_i)<S(M_\ell))
\Prob(M_\ell|R_a,M_i)\Prob(R_a),
\end{equation}
where $\Prob(R_a)$ and $\Prob(M_\ell|R_a,M_i)$ are same as in Eq.~\eqref{eq:transitionprobs}, and $\Prob(S(M_i)<S(M_\ell))$ is the probability that the current global model $M_i$ has a more favorable score than the proposed model $M_\ell$. The right hand side of Eq.~\eqref{eq:sticky} can be calculated for all $L$ models given $M_T$. This probability captures the tendency of scientific community to stay at a visited model. Calculated for $M_G^{(t+1)}=M_G^{(t)}=M_T$, it gives the stickiness of the true model.

When there are replicators in the model, we measure the tendency of the scientific community to maintain consensus on the true model by the proportion of time that true model remains as global model given that it was already the global model. Stickiness is the complement of the probability that a global model switches from true model to another model.

{\bf How long does scientific community stay on the true model?}
By theory of ergodic Markov chains, the probability that each model is selected as the global model converges to a constant value independent of the current global model $M_G^{(t)}$. The limiting probabilities of a model being selected as a global model reflect the long-term behavior of the scientific community and the proportion of time spent on each model. These limiting probabilities are given by $\lim_{t\rightarrow\infty}\mathbf{P}^t$.

When there are replicators in the model, we assess the mean time the scientific community spends on a model (and in particular the true model) as a consensus by the proportion of times that a model is selected as the global model. We use the proportion of times true model is selected as global model as a proxy for the limiting probabilities of selecting the true model as the global model in the process of scientific discovery with no replication.

{\bf How reproducible are the results of experiments?} Under ABM, we define the rate of reproducibility as the expectation of Bernoulli distributed random variable $\mathbf{I}_{\{M_G^{(t+1)}=M_G^{(t)}|R_o^{(t)}\}}$,
which is the indicator function that takes the value 1 if the global model at time $t+1$ is the same as the global model at $t$ given that a replication experiment is performed at time $t$. The parameter of the Bernoulli distribution is the probability of reproducibility
$\Prob(M_G^{(t+1)}=M_G^{(t)}|R_o^{(t)})$. Further, we define the rate of reproducibility when the true model is the global model by
$\mathbf{I}_{\{M_G^{(t+1)}=M_G^{(t)}|R_o^{(t)},M_G^{(t)}=M_T\}}$ which also determines the rate of reproducibility when the true model is not the global model
$\mathbf{I}_{\{M_G^{(t+1)}=M_G^{(t)}|R_o^{(t)},M_G^{(t)}\neq M_T\}}$. We estimate these rates of reproducibility by Monte Carlo integration of ABM simulations.
\subsubsection*{Monte Carlo estimates of rates of reproducibility}
\label{App:MonteCarloEstimate}
The random variable $\mathbf{I}_{\{M_G^{(t+1)}=M_G^{(t)}|R_o^{(t)}\}}$
which takes the value $1$ if the global model at time $t+1$ is equal to global model at time $t$ given that at time $t$ we have chosen a replicator, is a Bernoulli distributed random variable. Its mean is given by
$\E(\mathbf{I}_{\{M_G^{(t+1)}=M_G^{(t)}|R_o^{(t)}\}})$ whose Monte Carlo estimate is given by
\begin{equation}
\nonumber
\widehat{\E}(\mathbf{I}_{\{M_G^{(t+1)}=M_G^{(t)}|R_o^{(t)}\}})=\frac{1}{V}\sum_{v=1}^V
\mathbf{I}_{\{M_G^{(t+1)}=M_G^{(t)}|R_{o_v}^{(t)}\}},
\end{equation}
where $R_{o_v}^{(t)}$ is the $v$th instance a replicator is chosen. The rate of reproducibility when the true model is global model is estimated by
\begin{equation}
\nonumber
\widehat{\E}(\mathbf{I}_{\{M_G^{(t+1)}=M_G^{(t)}|R_0^{(t)},M_G^{(t)}=M_T\}})=\frac{1}{V_T}\sum_{v=1}^{V_T}
\mathbf{I}_{\{M_G^{(t+1)}=M_G^{(t)}|R_{0_v}^{(t)},M_G^{(t)}=M_T\}}
\end{equation}

which also implies
\begin{equation}
\nonumber
\widehat{\E}(\mathbf{I}_{\{M_G^{(t+1)}=M_G^{(t)}|R_0^{(t)},M_G^{(t)}\neq M_T\}})=\frac{1}{V_N}\sum_{v=1}^{V_N}
\mathbf{I}_{\{M_G^{(t+1)}=M_G^{(t)}|R_{0_v}^{(t)},M_G^{(t)}\neq M_T\}}.
\end{equation}
Here, $V_T$ is the number of times a replicator is selected when the true model is global model and $V_N$ is the number of times a replicator is selected when the true model is not global model, and $V_T+V_N=V$.
%
\paragraph*{S5 Appendix.}
\label{S5_Appendix}
{\bf Predictor variables, error variance, and sample size.}
We uniformly randomly generate the value of the $j$th factor at the $i$th level $x_{ij}$ on the set $\{1,2,\cdots, 100\}$ for all $i$ and $j$. We fix the sample size $n=100$ and calibrate the ratio of the error variance $\sigma^2$ to expected value of the model at the mean value of the predictors $\E(y|\mu_x)$, where $\mu_x=\E(x)$. We use a standardized linear regression model so that if the $i$th observed response is $y_i^*$, then we set $y_i=(y_i^*-\bar{y}^*)/s^*$, where $\bar{y}^*$ is the sample mean and $s^*$ is the sample standard deviation of $y_i^*,\;i=1,2,\cdots, n$. Implementing standardized linear regression model allows us to precisely specify $\sigma^2:\E(y|\mu_x)$. We fix three levels for $\sigma^2:\E(y|\mu_x)$. $(1:4)$, $(1:1)$, and $(4:1)$.
We set the true regression coefficients to be Dirichlet distributed with unit parameters or equivalently, uniformly distributed on the interval $(0,1)$ with the constraint that they sum to 1. Thus, all regression coefficients have the same mean effect size. We set the correlation between the first factor $x_1$ and other factors to a small value of $0.2$ to avoid orthogonality between predictors which is an idealized case that we believe rarely achievable in practice. We performed additional analyses with higher correlation between predictor variables and found that correlation between predictors does not affect the results in our system unless it is extremely high and causes multicollinearity.

{\bf Model selection criteria.}
SC is defined by $p\log(n)-\log \Prob(D|\hat{\theta},M)$, where $p$ is the number of model parameters, $n$ is the sample size, and $\hat{\theta}$ is the maximum likelihood estimate of model parameters under $M$. Akaike's Information Criterion (AIC) is defined by $2p-\log \Prob(D|\hat{\theta},M)$. For both SC and AIC, a smaller value indicates a better model performance than a larger value. The maximized likelihood $\log \Prob(D|\hat{\theta},M)$ rewards model fit equally in SC and AIC. The first term in these formulas penalize the model complexity, with SC penalizing complex models more heavily for $n\geq8$ (with $\log(n)$) than AIC as the sample size increases. SC has the desirable property that when the true model generating the data is in the universe of candidate models, it selects the true model with probability $1$ as $n\rightarrow \infty$, in other words it is consistent. For model $M$ fitted to $D$, we calculate the model selection criteria SC and AIC in a computationally efficient way.

{\bf Description of scientists and scientist populations.}
We let $\R=\{R_{Rey}, R_{Tess}, R_{Mave}, R_{Bo}\}$ and define the transition probabilities for these strategies as follows. For $R_{Rey}$,
\[ \Prob(M_P^{(t)}=M_j| M_P^{(t-1)}=M_i,M_G^{(t)}=M_G^{(t-1)}) =
  \begin{cases}
    1      & \quad \text{if } i=j,\\
    0      & \quad \text{otherwise.}
    \end{cases}
\]
For $R_{Tess}$,
\[ \Prob(M_P^{(t)}=M_i|M_G^{(t)}) =
  \begin{cases}
    1/(m+1)      & \quad \text{if } M_i \text{ is one of } m \text{ models}\\                  & \quad \text{one term away from } M_G^{(t)},\\
    m_o & \quad \text{otherwise.}
    \end{cases}
\]
If $R_{Rey} \in \R$  then $m_o=0$ and  $m_o=1/[(L-m)(m+1)]$ otherwise.
For $R_{Mave},\; \Prob(M_P^{(t)}=M_i|M_G^{(t)}) =1/L,\;
i=1,2,\cdots,L$. For $R_{Bo}$,
\[ \Prob(M_P^{(t)}=M_i|M_G^{(t)}) =
  \begin{cases}
    1/(m+1)      & \quad \text{if } M_i \text{ is one of } m \text { models one}\\
    & \quad \text{interaction larger than } M_G^{(t)},\\
    m_o & \quad \text{otherwise.}
    \end{cases}
\]
If $R_{Rey} \in \R$  then $m_o=0$ and  $m_o=1/[(L-m)(m+1)]$ otherwise.
\begin{table}[H]
\captionsetup{labelformat=empty}
\caption{{\bf S6 Table. Populations of scientists with varying proportions of scientist types.}} \label{S6_Table}
\begin{center}
\begin{tabular}{ccccccc}
\hline
\hline
Scenario & Population & Rey  & Mave & Tess & Bo \\
\hline
\rule{0pt}{3ex}
\multirow{4}{*}{Without Replicator} &  Tess Dominant & 0 & 0.005 & 0.99 & 0.005\\
                  & Mave Dominant & 0 & 0.99 & 0.005 & 0.005\\
                  & Bo Dominant & 0 & 0.005 & 0.005 & 0.99\\
                  & All Equal & 0 & $0.\bar{3}$& $0.\bar{3}$ & $0.\bar{3}$\\
\rule{0pt}{3ex}
\multirow{4}{*}{With Replicator}    & Rey Dominant & 0.99 & $0.00\bar{3}$ & $0.00\bar{3}$ & $0.00\bar{3}$\\
                  & Tess Dominant  &$ 0.00\bar{3}$ & $0.00\bar{3}$ &0.99& $0.00\bar{3}$\\
                                    & Mave Dominant &$ 0.00\bar{3}$ & 0.99 & $0.00\bar{3}$ & $0.00\bar{3}$\\
                  & Bo Dominant &$ 0.00\bar{3} $& $0.00\bar{3}$ & $0.00\bar{3}$& 0.99\\
                  & All Equal & 0.25 & 0.25 & 0.25 & 0.25\\
\hline
\hline
\end{tabular}
\end{center}
\end{table}

\paragraph*{S7 Code and Data.}
\label{S7_CodeAndData}
The code to perform the simulations and analyze the data generated in this project, and a summary data set are included as a Git repository at \url{https://github.com/gnardin/CRUST}. Refer to the description in the main page of this repository for further instructions and details of the implementation.

\paragraph*{S8 Appendix.}
\label{S8_Appendix}
{\bf Monte Carlo Estimates of Model Comparisons.}
Let $\mathbf{y}$ denote $n \times 1$ vector of responses generated by the true model and $\mathbf{X}$ be $n \times p$ matrix of predictors, where $p$ is the number of parameters in the fitted model. Under the Gauss-Markov assumptions $\E(\epsilon_i)=0, \;Var(\epsilon_i)=\sigma^2, \;Cov(\epsilon_i,\epsilon_j)$ for all $i,j$ we denote the vector of joint maximum likelihood estimates for the regression coefficients by $\bm{\hat{\beta}}$. By definition of Schwarz Criterion
\begin{equation*}
S(M)=2p\log(n)-2\log\Prob(\y|\X,\bm{\hat{\beta}}).
\end{equation*}
If Akaike's Information Criterion is used, $2p$ replaces $2p\log(n)$.
The loglikelihood in the second term is equal to $n$ times the log of the residual sums of squares and it can be written as
\begin{equation*}
\log\Prob(\y|\X,\bm{\hat{\beta}})=n\log( \y^\prime \y -\y^\prime\X(\X^\prime\X)^{-1}\X^\prime\y)+C,
\end{equation*}
where $C$ is a term dependent only on $M_T$. For transition probabilities, we are interested in $\Prob(S(M_i)<S(M_j))$. We have
\begin{eqnarray}\label{eq:DeltaSchwarz}
\nonumber
S(M_i)-S(M_j)&=&(p_i-p_j)\log(n)+n\log(\y^\prime\X_i(\X_i^\prime \X_i)^{-1}\X_i^\prime\y)\\
& &-n\log(\y^\prime \X_j (\X_j^\prime \X_j)^{-1} \X_j^\prime \y),
\end{eqnarray}
where subscripts $i,j$ now denote quantities that depend on model $M_i$ and $M_j$.
The random variable $\mathbf{I}_{\{S(M_i)-S(M_j)<0|M_T\}}$ is Bernoulli distributed with probability of success $\Prob(S(M_i)-S(M_j)<0)$ which is equal to its expectation $\E(\mathbf{I}_{\{S(M_i)-S(M_j)<0|M_T\}})$ whose Monte Carlo estimate is given by
\begin{equation}\label{ap:eq:MonteCarlo}
\widehat{\E}(\mathbf{I}_{\{S(M_i)-S(M_j)<0|M_T\}})=\frac{1}{V}\sum_{v=1}^V
\mathbf{I}_{\{S(M_i)-S(M_j)<0|\y_v\}}.
\end{equation}
An estimate of $\Prob(S(M_i)-S(M_j)<0)$ is obtained using Eq.~\eqref{ap:eq:MonteCarlo} conditional on true model $M_T$ and its predictors $X_T$ as follows. First, generate the set of $k$ predictor variables and build $\X_i$ and $\X_j$ for $M_i$ and $M_j$ respectively. Then generate $\bm{\beta_{T_v}},\;v=1,2,\cdots, V$ independently of each other. Finally, simulate $\y_v|\X_T,\bm{\beta_{T_v}}$ from the normal distribution with expected value $\E(\y_v)=\X_T\bm{\beta_{T_v}}$ and variance $\sigma^2$.  Each realization $(\y_1,\y_2,\cdots,\y_v)$ is used in Eq.~\eqref{eq:DeltaSchwarz} to assess $S(M_i)-S(M_j)<0$ and the estimate is obtained using the mean in Eq.~\eqref{ap:eq:MonteCarlo}.

\begin{table}[H]
\captionsetup{labelformat=empty}
\caption{{\bf S9 Table. Parameter values we used for the completely randomized factorial design experiment to generate the results reported in this work.}}
\label{S9_Table}
  \begin{center}
  \begin{tabular}{ll}
    \toprule\noalign{\smallskip}
    \textbf{Parameter}  & \textbf{Value}\\
    \midrule\noalign{\smallskip}
    \texttt{replications} & 100\\
    \hline\noalign{\smallskip}
    \texttt{timesteps}    & 11000\\
    \hline\noalign{\smallskip}
    \texttt{k}            & 3\\
    \hline\noalign{\smallskip}
    \texttt{sigma}        & 0.2, 0.5 and 0.8\\
    \hline\noalign{\smallskip}
    \texttt{sampleSize}   & 100\\
    \hline\noalign{\smallskip}
    \texttt{trueModel}    & \texttt{x1 + x2},\\
                          & \texttt{x1 + x2 + x3 + x1x2}, and\\
                          & \texttt{x1 + x2 + x3 + x1x2 + x1x3 + x2x3}\\
    \hline\noalign{\smallskip}
    \texttt{correlation}  & 0.2\\
    \hline\noalign{\smallskip}
    \texttt{nRey}         & 1 and 300\\
    \hline\noalign{\smallskip}
    \texttt{nTess}        & 1 and 300\\
    \hline\noalign{\smallskip}
    \texttt{nBo}          & 1 and 300\\
    \hline\noalign{\smallskip}
    \texttt{nMave}        & 1 and 300\\
    \hline\noalign{\smallskip}
    \texttt{modelCompare} & \texttt{AIC} and \texttt{BIC}\\
    \hline\noalign{\smallskip}
    \texttt{ndec}         & 4\\
    \bottomrule\noalign{\smallskip}
  \end{tabular}
  \end{center}
\end{table}

\begin{figure}[H]
\includegraphics[width=1\textwidth]{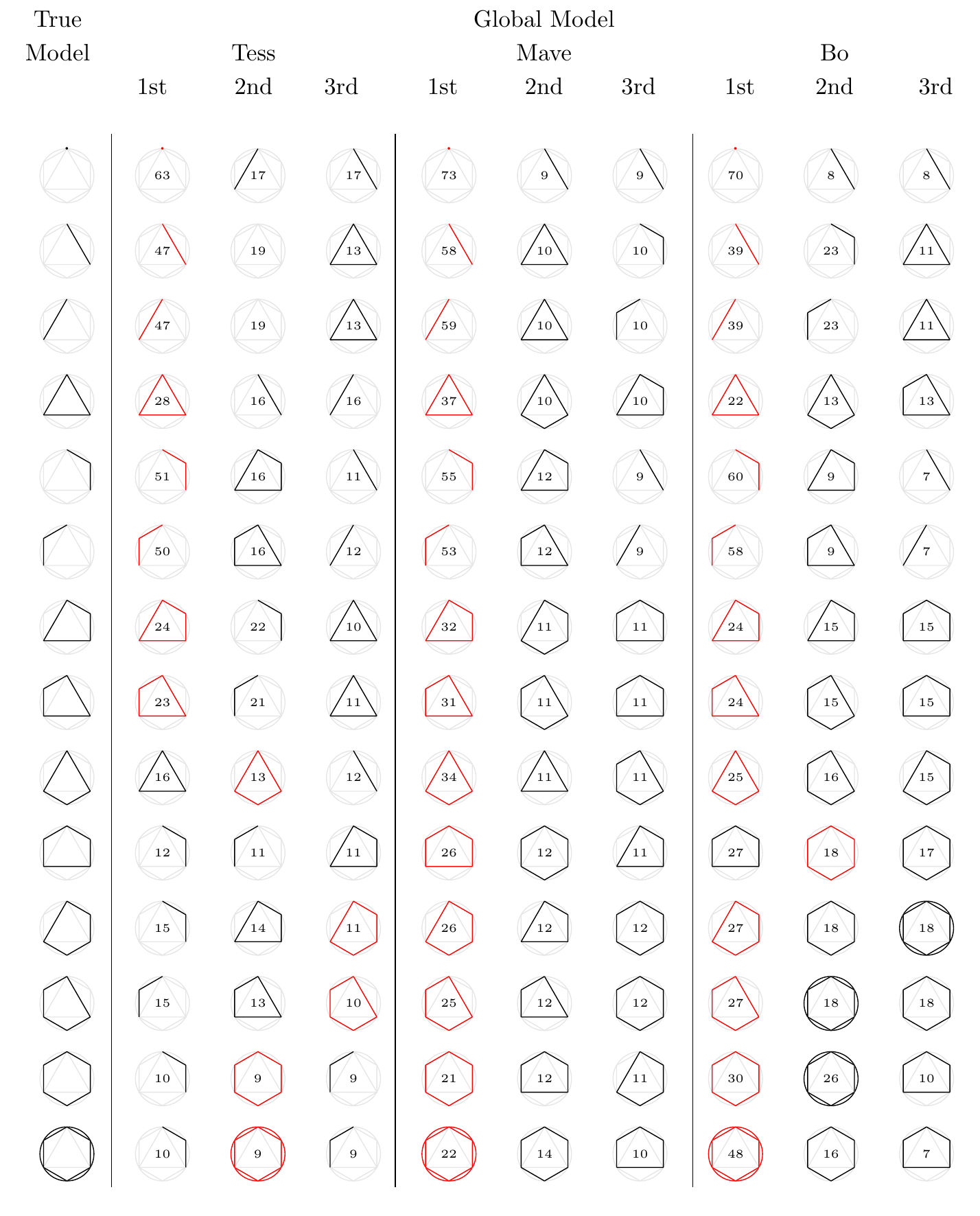}
\captionsetup{labelformat=empty}
\caption{{\bf S14 Figure. The three most visited models for scientist populations with one dominant type and the proportion of time spent at each true model, when SC is the model comparison statistic and noise equals signal in a system with no replication.} For $\sigma^2:\E(y|\mu_x)=(1:1)$, proportion of time spent by a model as the global given a true model, assessed by SC. Three most visited models are shown. Numbers show time spent at each model in percent points. True models are in red. Under SC and with this level of noise, \textit{Tess-}dominant population performs more poorly than both \textit{Mave-} and \textit{Bo-}dominant populations, spending much less time in the true model.}
\label{S14_Figure}
\end{figure}

\begin{figure}[H]
\includegraphics[width=1\textwidth]{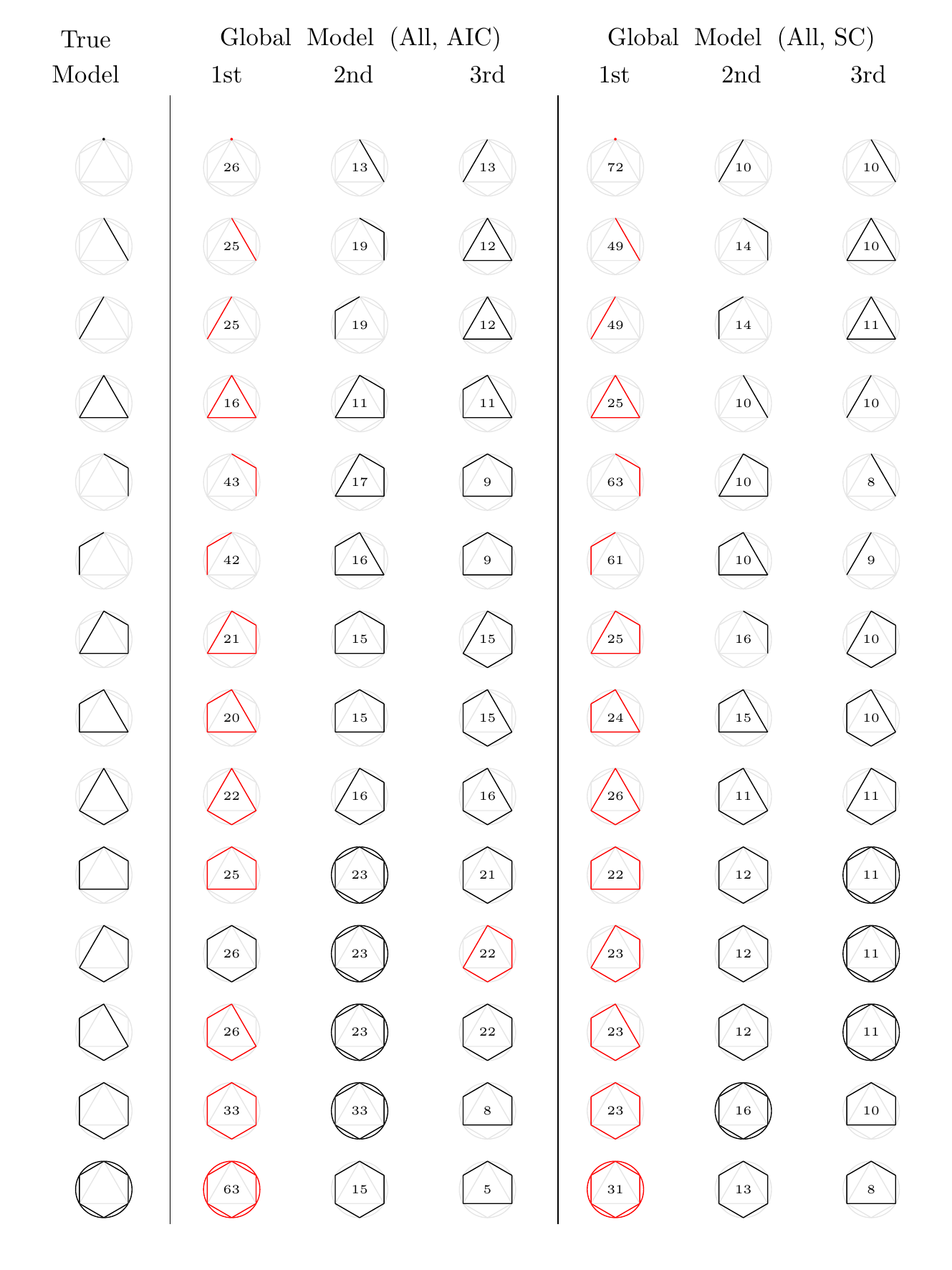}
\captionsetup{labelformat=empty}
\caption{{\bf S15 Figure. The three most visited models by the epistemically diverse population for each true model and when noise equals signal in a system with no replication.} For $\sigma^2:\E(y|\mu_x)=(1:1)$, proportion of time spent by a model as the global given a true model for an epistemically diverse population. Three most visited models are shown for AIC and SC. Numbers show time spent at each model in percent points. True models are in red. For epistemically diverse population, the true model is the most visited model, for all true models except one under AIC and for all true models under SC. It spends more time in simpler models under SC than under AIC.}
\label{S15_Figure}
\end{figure}

\begin{figure}[H]
\includegraphics[width=1\textwidth]{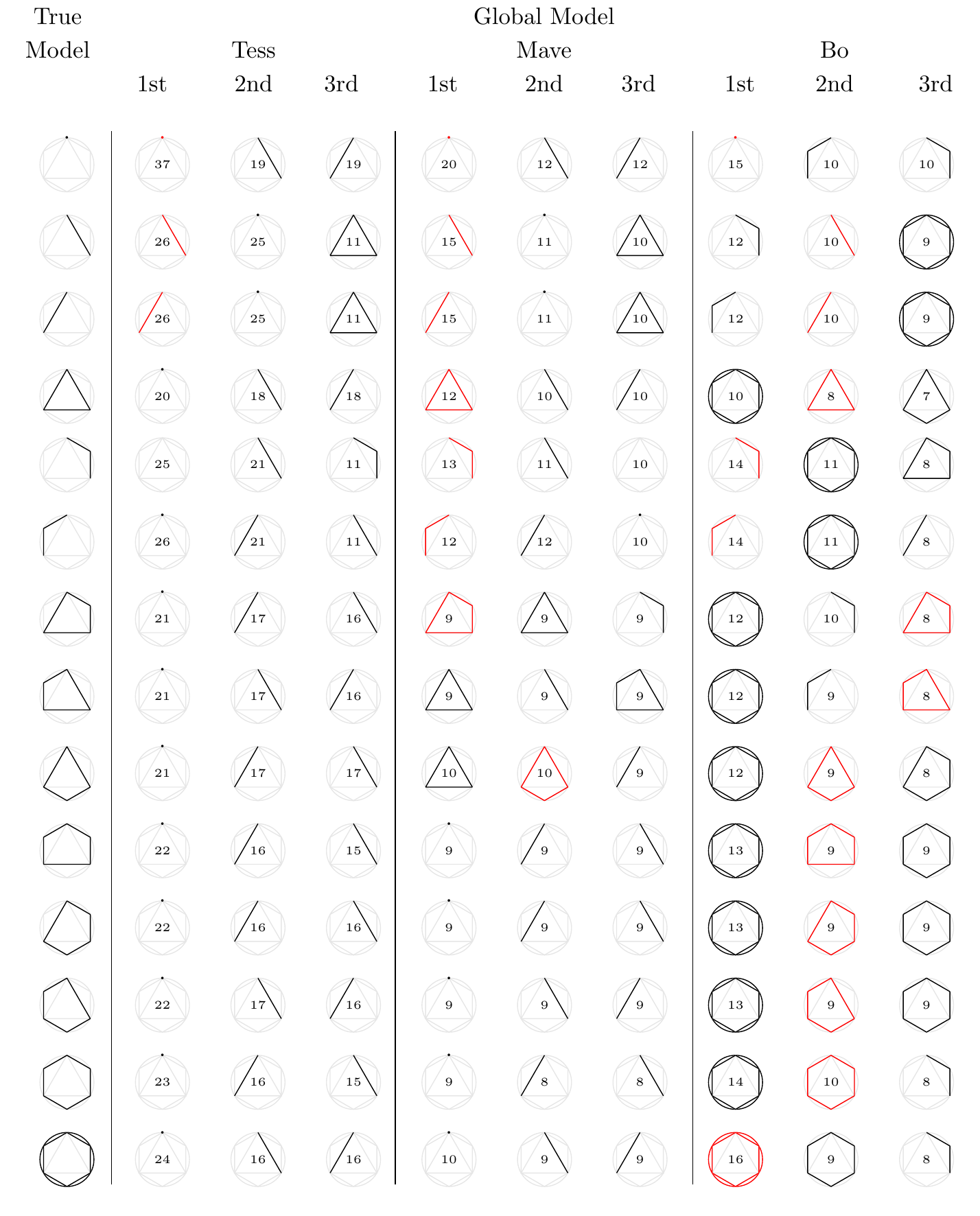}
\captionsetup{labelformat=empty}
\caption{{\bf S16 Figure. The three most visited models for scientist populations with one dominant type and the proportion of time spent at each true model, when AIC is the model comparison statistic and noise-to-signal ratio is $4:1$ in a system with no replication.} For $\sigma^2:\E(y|\mu_x)=(4:1)$, proportion of time spent by a model as the global given a true model, assessed by AIC. Three most visited models are shown. Numbers show time spent at each model in percent points. True models are in red. When the level of noise in the system is extremely high, all heterogeneous populations but \textit{Bo-}dominant fail to capture the true model for many true models and spend little time at it overall. For \textit{Bo-}dominant population, true model is among top three most visited models across all true models.}
\label{S16_Figure}
\end{figure}

\begin{figure}[H]
\includegraphics[width=1\textwidth]{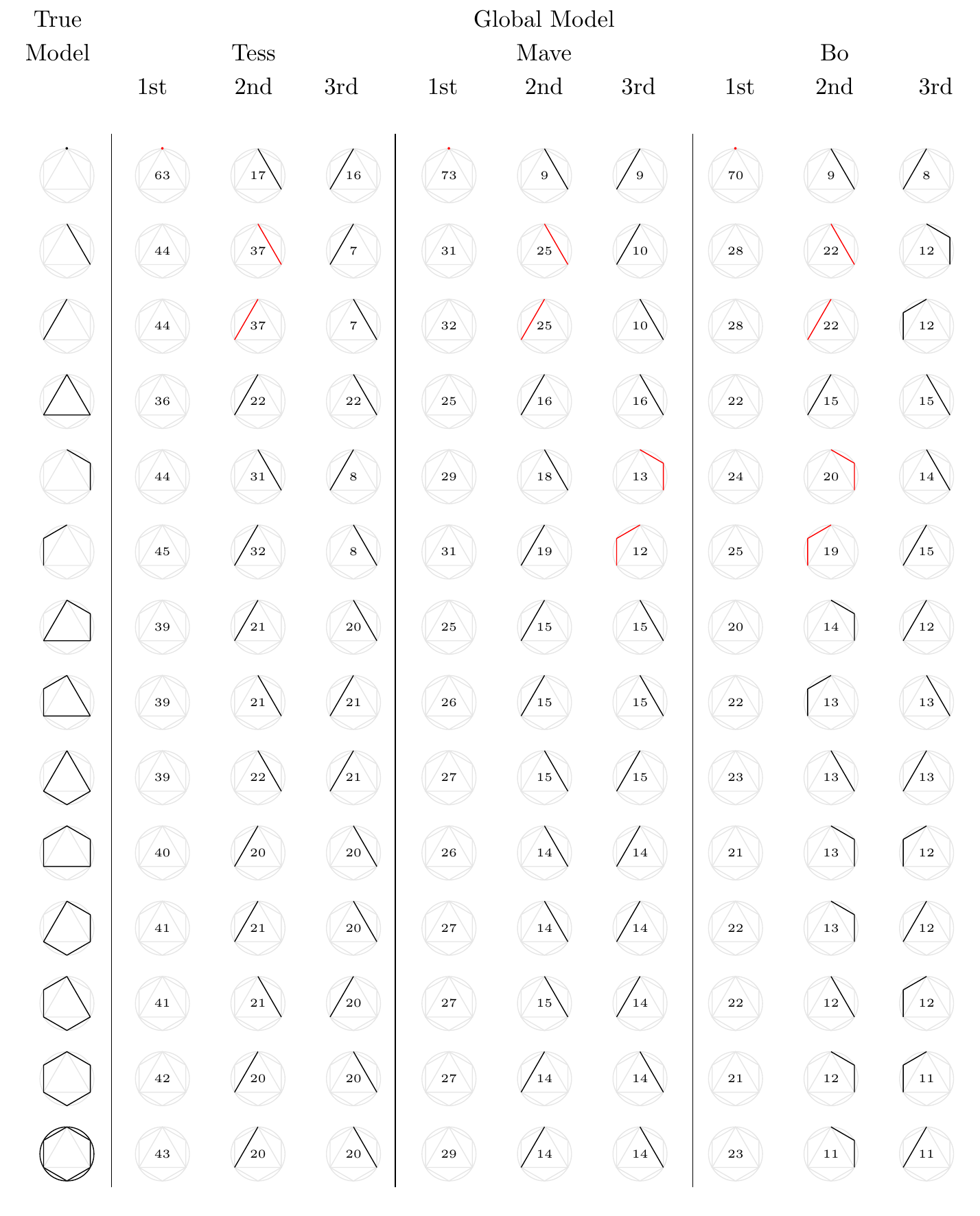}
\captionsetup{labelformat=empty}
\caption{{\bf S17 Figure. The three most visited models for scientist populations with one dominant type and the proportion of time spent at each true model, when SC is the model comparison statistic and noise-to-signal ratio is $4:1$ in a system with no replication.} For $\sigma^2:\E(y|\mu_x)=(4:1)$, proportion of time spent by a model as the global given a true model, assessed by SC. Three most visited models are shown. Numbers show time spent at each model in percent points. True models are in red. Under SC, the high performance of \textit{Bo-}dominant population is dampened and all homogeneous populations perform very poorly.}
\label{S17_Figure}
\end{figure}

\begin{figure}[H]
\includegraphics[width=1\textwidth]{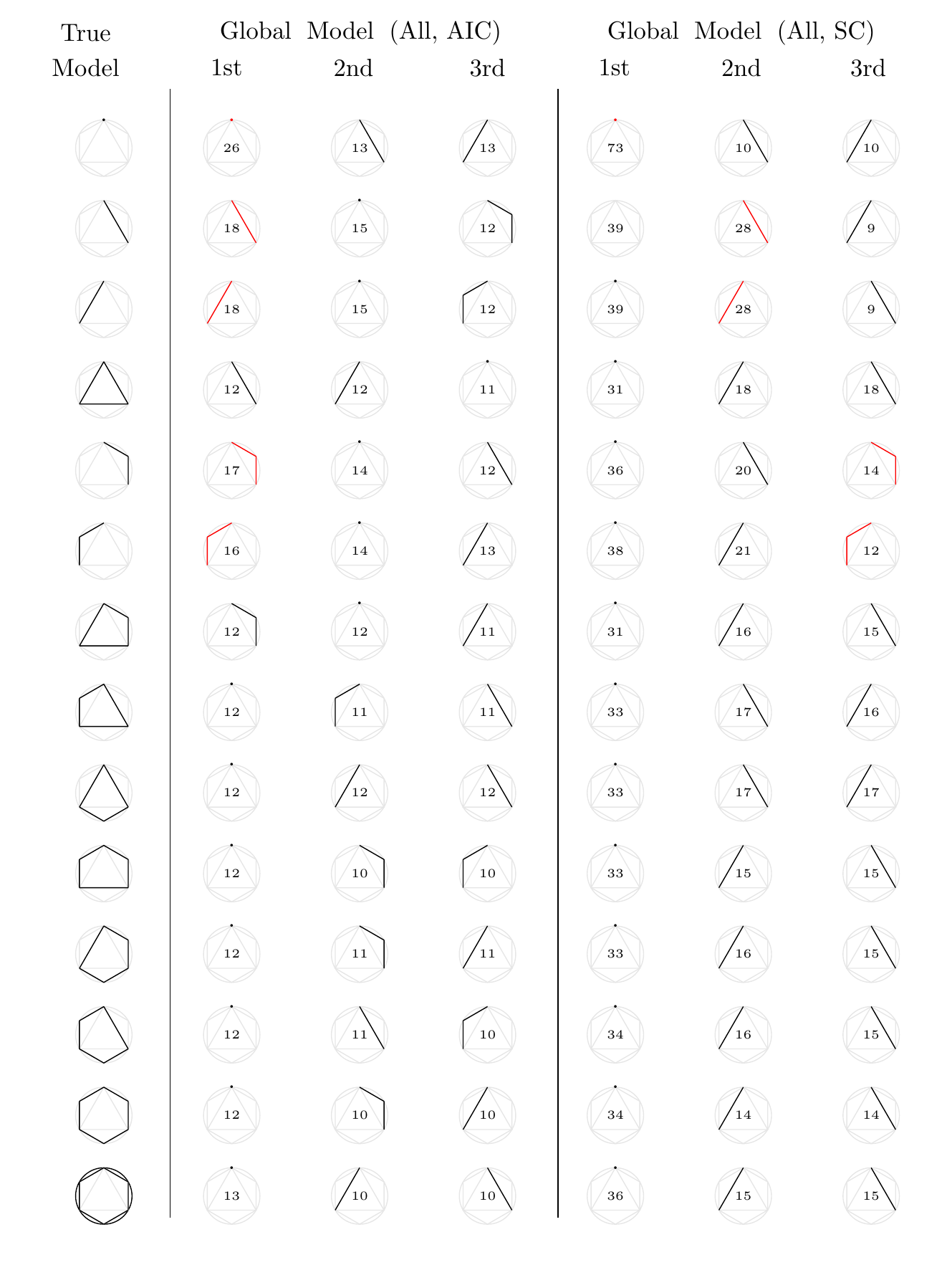}
\captionsetup{labelformat=empty}
\caption{{\bf S18 Figure. The three most visited models by the epistemically diverse population for each true model and noise-to-signal ratio is $4:1$ in a system with no replication.} For $\sigma^2:\E(y|\mu_x)=(4:1)$, proportion of time spent by a model as the global given a true model for an epistemically diverse population. Three most visited models are shown for AIC and SC. Numbers show time spent at each model in percent points. True models are in red. When the system noise is high, even the epistemically diverse population cannot prevent poor performance. True model is not captured for most models and most of the time, under both model comparison statistics.}
\label{S18_Figure}
\end{figure}

\begin{figure}[H]
\includegraphics[width=1\textwidth]{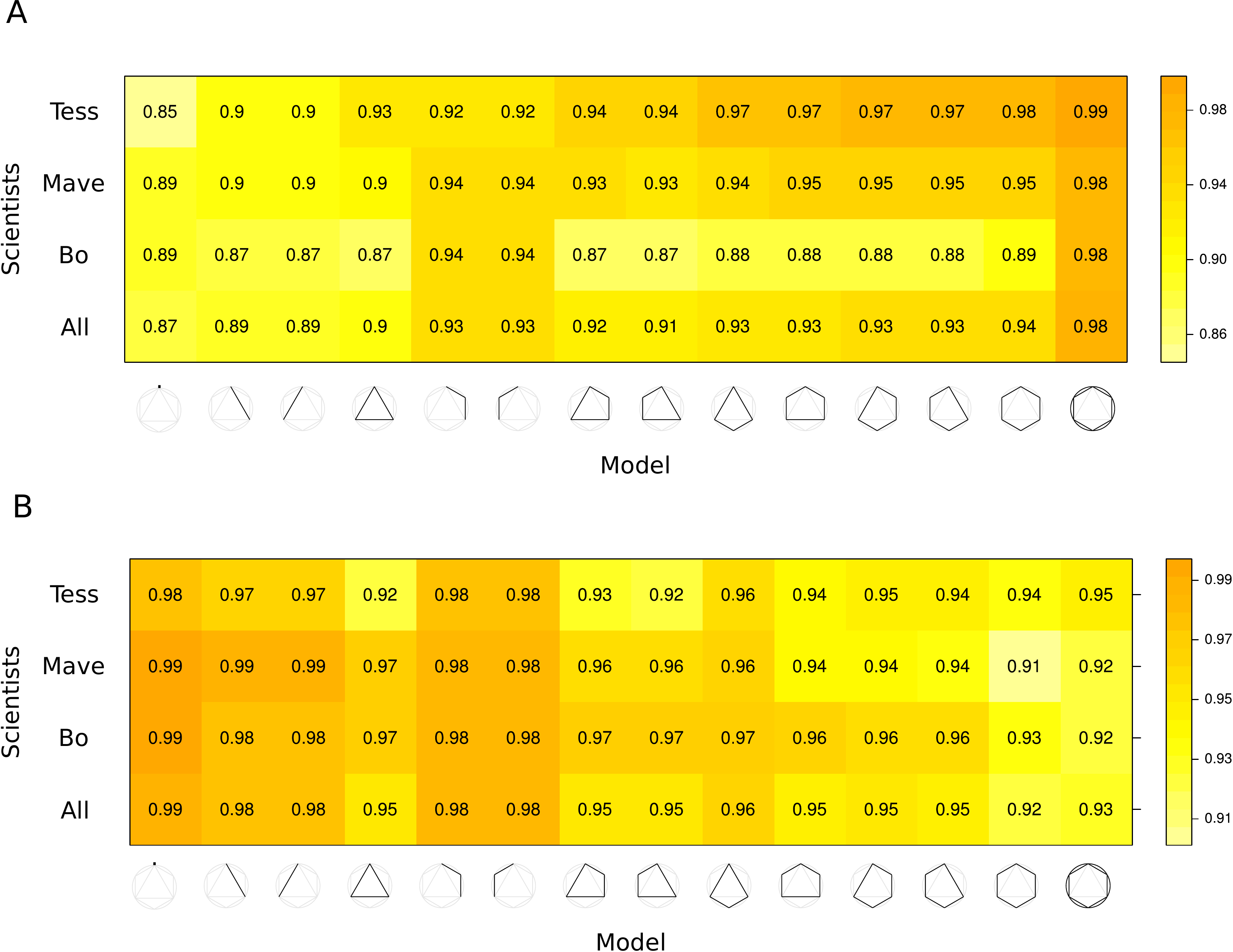}
\captionsetup{labelformat=empty}
\caption{{\bf S19 Figure. The true model stickiness when noise-to-signal ratio is $1:1$ in a system with no replication.} For $\sigma^2:\E(y|\mu_x)=(1:1)$, stickiness of each true model as global model for each scientific population (vertical axis) for AIC (A) and SC (B). The true model is still sticky when noise is set to be equal to the signal in a system with no replication. True model is stickiest for \textit{Tess}-dominant population (increasing with complexity) and least sticky for \textit{Bo}-dominant population under AIC. Under SC, true model stickiness is even higher, and all populations perform comparably well.}
\label{S19_Figure}
\end{figure}

\begin{figure}[H]
\includegraphics[width=1\textwidth]{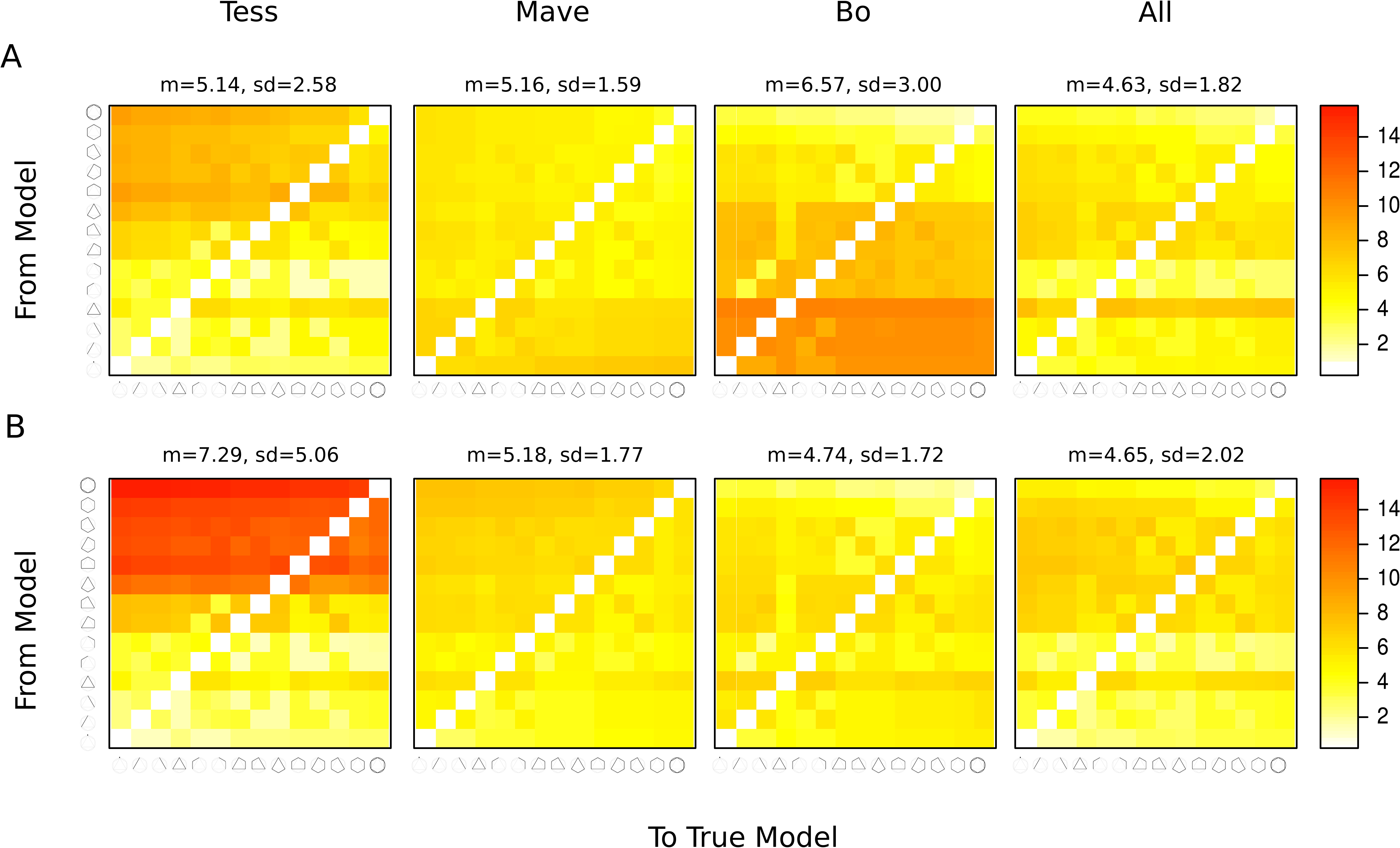}
\captionsetup{labelformat=empty}
\caption{{\bf S20 Figure. The mean first passage time to true model when noise is set to be equal to the signal in a system with no replication.} For $\sigma^2:\E(y|\mu_x)=(1:1)$, the mean first passage time from each initial model (vertical axis) to each true model (horizontal axis) using AIC (A) and SC (B) as model comparison statistics per scientist populations. All means epistemically diverse; all others dominant in one type. Epistemically diverse population reaches truth fastest under both AIC and SC. Interestingly, under AIC \textit{Bo}-dominant population is the slowest to reach the truth whereas under SC, it is the \textit{Tess}-dominant population, especially when starting from complex initial models.}
\label{S20_Figure}
\end{figure}

\begin{figure}[H]
\includegraphics[width=1\textwidth]{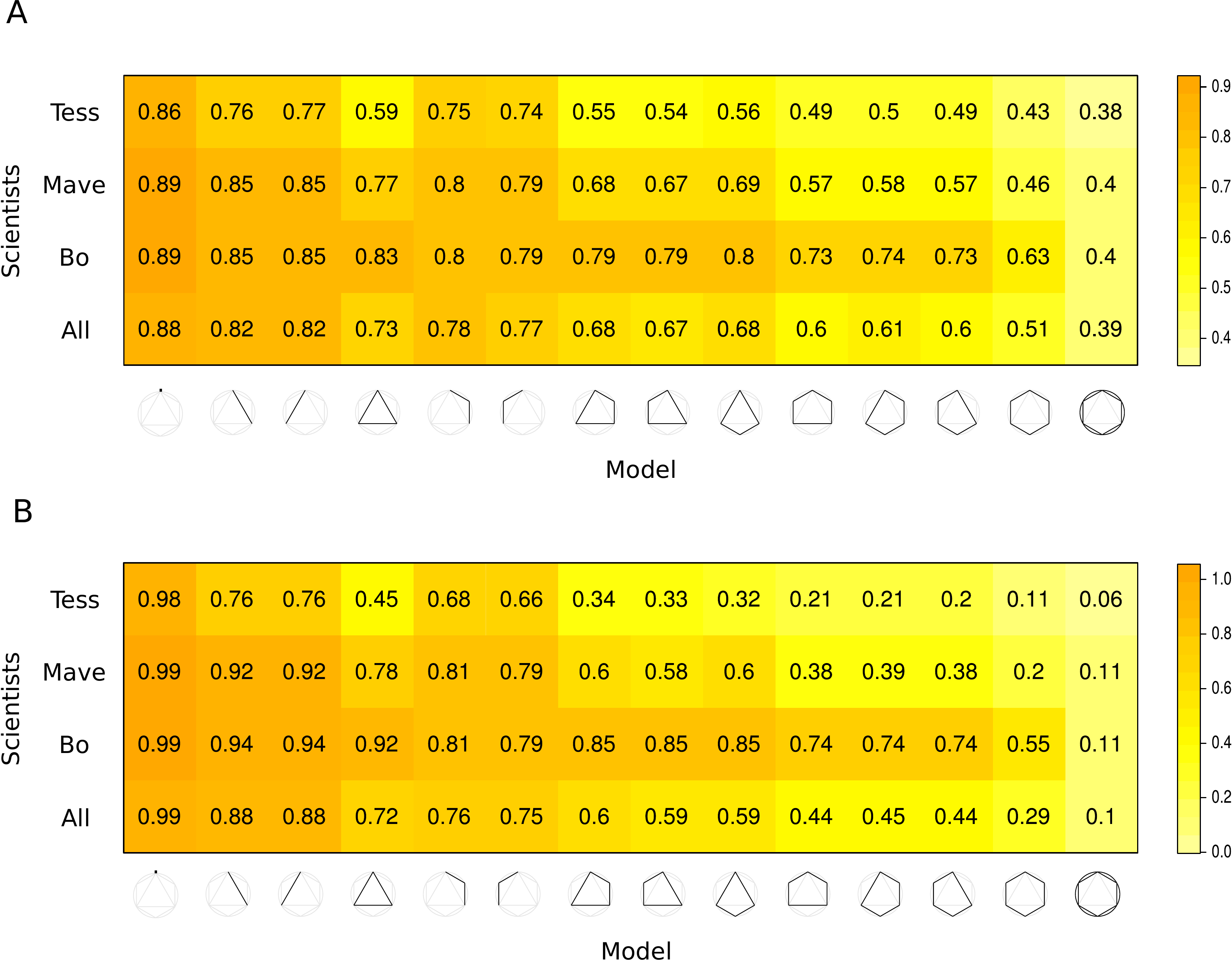}
\captionsetup{labelformat=empty}
\caption{{\bf S21 Figure. The true model stickiness when noise-to-signal ratio is $4:1$ in a system with no replication.} For $\sigma^2:\E(y|\mu_x)=(4:1)$, stickiness of each true model as global model for each scientific population (vertical axis) for AIC (A) and SC (B). In this scenario, we observe a substantial decrease in true model stickiness, especially for complex models, both under AIC and SC. Level of noise in the system appears to have a large effect on whether true model will stay as global model once it is hit. In such cases, \textit{Bo}-dominant population appears to perform better than other populations but still not as well as the cases with lower noise.}
\label{S21_Figure}
\end{figure}

\paragraph{S23 Appendix.}
\label{S23_Appendix} \textbf{Reproducibility does not imply discovery of truth.}

The rate of reproducibility is the probability of the global model staying the same model after a replication experiment.
In this section, we show that the rate of reproducibility has no effect on other desirable properties of scientific discovery including: the probability that a model is selected as the global model in the long run, the mean first time to hit a model, and stickiness of a model.
The key observation for the result is that all of these desirable properties of scientific discovery are properties of the stochastic process of scientific discovery defined by the Markov chain. A Markov chain is characterized solely by its transition probability matrix. Below, we show that the rate of reproducibility, does not affect the transition probability matrix in our model. Therefore, any property of the Markov chain is not affected by the rate of reproducibility. To keep the notation tractable and without loss of generality we assume that \textit{Rey}, the replicator is not chosen consecutively in the process. Generalization to the case where \textit{Rey} is chosen consecutively is by induction and using Eq.~\eqref{eq:1}.

When \textit{Rey}, the replicator, is in the system, the model is a second order Markov chain, which has transitions specified over two time steps. We let the triplet of indices $(i,j,\ell)$ to be associated with a transition in this second order Markov chain, where $i$ is the beginning state and $\ell$ is the final transition state through state $j$. We let $R_{Rey}$ to be the replicator strategy and $R_{Rey^\prime}$ to be any non-replicator strategy. $P(R_{Rey}) = 1-P(R_{Rey^\prime})$ is the probability that the second step is a replication experiment. The probability that the global model transitions to $M_\ell$ in the second step given that $M_i$ was the beginning global model can be written as
\begin{equation}\label{transprob}
P(R_{Rey^\prime})\sum_{j=1}^{L}p_{ij\ell}+P(R_{Rey})\sum_{j=1}^{L}q_{ij\ell},
\end{equation}
where $p_{ij\ell}$ and $q_{ij\ell}$ are transition probabilities when a replicator is not selected and is selected, respectively. The first term in Eq.~\eqref{transprob} does not involve a replicator so to establish whether the rate of reproducibility affects transition probabilities it is sufficient to focus on the second term.
There are only two nonzero terms in $\sum_{j=1}^{L}q_{ij\ell}$. The first term is
$$q_{ii\ell}=P(S(M_\ell)<S(M_i))P(S(M_\ell)<S(M_i))P(M_\ell|M_i),$$
where on the right hand side: $M_\ell$ is proposed with probability $P(M_\ell|M_i)$ and wins against $M_i$ with probability $P(S(M_\ell)<S(M_i))$
at the first step, and in the replication (second step) it wins again with probability $P(S(M_\ell)<S(M_i))$ to stay as the global model.
The second term nonzero term is
$$q_{i\ell\ell}=P(S(M_\ell)<S(M_i))P(S(M_\ell)>S(M_i))P(M_\ell|M_i),$$
where on the right hand side: $M_\ell$ is proposed with probability $P(M_\ell|M_i)$ and loses against $M_i$ with probability $P(S(M_\ell)>S(M_i))$ at the first step, and in the replication (second step) it wins against $M_i$ with probability $P(S(M_\ell)<S(M_i))$ to become the global model. Here, $q_{ii\ell}$ and $q_{i\ell\ell}$ are the probabilities of reproducing and not reproducing a result. Further, only their sum contributes to the Markov transition probability matrix (Eq.~\eqref{transprob}). We write
\begin{align} \nonumber
\sum_{j=1}^{L}q_{ij\ell}&=q_{ii\ell}+q_{i\ell\ell}\\ \nonumber
&=P(S(M_\ell)<S(M_i))P(S(M_\ell)<S(M_i))P(M_\ell|M_i)\\ \nonumber
&+P(S(M_\ell)<S(M_i))P(S(M_\ell)>S(M_i))P(M_\ell|M_i)\\ \nonumber
&=P(S(M_\ell)<S(M_i))P(M_\ell|M_i)[P(S(M_\ell)<S(M_i))+P(S(M_\ell)>S(M_i))],
\end{align}
where the first and second term within brackets determine whether a result is reproduced or not reproduced, respectively, and sum to $1$. Therefore we have
\begin{equation} \label{transprob2}
\sum_{j=1}^{L}q_{ij\ell}
=P(S(M_\ell)<S(M_i))P(M_\ell|M_i),
\end{equation}
which shows that whether a result is reproduced has no bearing on the calculation of transition probability from $M_i$ to $M_\ell$. Since the transition probability matrix characterizes a Markov chain, the rate of reproducibility cannot affect other desirable properties of scientific discovery including: the probability that a model is selected as the global model in the long run, the mean first time to hit a model, and the stickiness of a model.

On the other hand, we do not claim that there cannot be correlation between desirable properties of scientific discovery and the rate of reproducibility. In fact, whether there is correlation depends on the strategies and their frequency in the population, and this also can be seen from our mathematical result. For transition to $M_\ell$ the rate of reproducibility is proportional to $\sum_{i\neq\ell}P(S(M_\ell)<S(M_i))$ and the transition probability to $M_\ell$ is proportional to $P(S(M_\ell)<S(M_i))$. If the effect of the first term $P(R_{Rey^\prime})\sum_{j=1}^{L}p_{ij\ell}$ in Eq.~\eqref{transprob} and the second term $P(M_\ell|M_i)$ in Eq.~\ref{transprob2} are small then there might be considerable correlation between desirable properties of scientific discovery and the rate of reproducibility. This relationship depends on a number of factors including the exact strategies of scientists in the population and the frequency of these scientists, and is very complex.
\paragraph*{S24 Appendix.} \label{S24_Appendix}
{\bf Correlations per scientist population.}
\begin{table}[!htbp]
\caption{Spearman rank-order correlation coefficients between \textit{rate of reproducibility} and other desirable properties of scientific discovery for each scientist population. \textit{Overall} is averaged over all scientist populations.}
\begin{center}
\begin{tabular}{ccccccc}
\hline
\hline
\textbf{$r_{SR}$} & Overall & Rey & Tess  & Mave & Bo & All equal  \\
\hline
\rule{0pt}{3ex}
Time spent at  &  -0.02 & 0.41 & 0.35 & 0.69 & -0.06 & 0.59\\
true model& & & & & & \\
Mean first passage &  0.26 & 0.03 & 0.023 & -0.13 & 0.03 & -0.06\\
time to true model& & & & & & \\
Stickiness &  0.55 & 0.74 & 0.87 & 0.76 & -0.03 & 0.82\\
\hline
\hline
\end{tabular}
\end{center}
\label{S24_Table}
\end{table}

\paragraph*{S26 Appendix.}\label{S26_Appendix}
{\bf ABM with soft research strategies.}
\begin{table}[!htbp]
\caption{Median and IQR of the mean (over $100$ replications for each parameter set) first passage time to true model calculated over runs with different model parameters for ABM with soft research strategies.}
\begin{center}
\begin{tabular}{cccccc}
\hline
\hline
 & Rey & Tess  & Mave & Bo & All equal \\
\hline
\rule{0pt}{3ex}
Median &  815 & 80 & 16 & 18 & 25\\
IQR &  1686 & 620 & 36.25 & 40 & 46\\
\hline
\hline
\end{tabular}
\end{center}
\label{S26_Table}
\end{table}

Mean first passage time to true model is faster for \textit{Tess-} and \textit{Bo-}dominant populations when scientists in the ABM simulations pursue soft (vs. hard) research strategies, consistent with results from Markov chain without replicator. The median of the mean first passage time for \textit{Mave-}dominant population is the same as the median we reported for ABM with hard research strategies because the maverick strategy has a well connected transition matrix, leading to quick discovery of the true model. Further, the representation of \textit{Mave}s in the epistemically diverse population makes this population also well connected, in ABM with both hard and soft research strategies. \textit{Rey-}dominant populations are the slowest to discover the truth because when replicators comprise $99\%$ of the population, there are not many new models being proposed and same experiments are repeated many times slowing down the discovery of the true model.
%

\begin{figure}[H]
\centering
\includegraphics[width=1\textwidth]{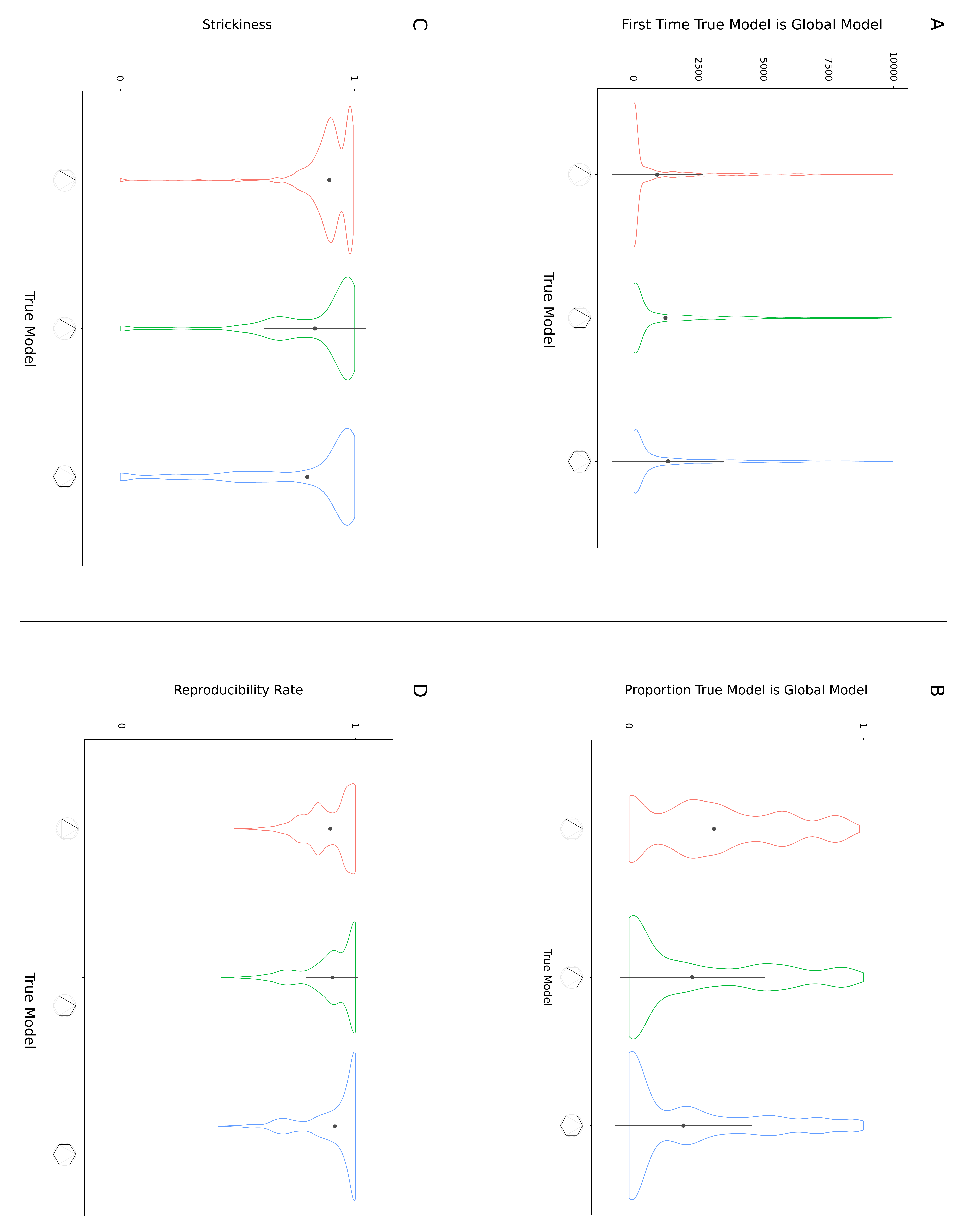}
\captionsetup{labelformat=empty}
\caption{{\bf S28 Figure. All properties of scientific discovery as a function of true model complexity in a system with replication.} Violin plots showing marginal effects of true model complexity on first mean passage time to true model (A), proportion of times the true model is global model (B), stickiness (C), and the rate of reproducibility (D). Complexity does not appear to have a substantial direct effect on any property and most of its effect comes through interactions with other model parameters.}
\label{S28_Figure}
\end{figure}

\section*{Acknowledgments}
This research made use of the resources of the High Performance Computing Center at Idaho National Laboratory, which is supported by the Office of Nuclear Energy of the US DoE under Contract No. DE-AC07-05ID14517. The funders had no role in study design, data collection and analysis, decision to publish, or preparation of the manuscript. 

%
%
%


\begin{thebibliography}{10}

\bibitem{Ramsey1931}
Ramsey FP.
\newblock Truth and probability (1926).
\newblock In: The Foundations of Mathematics and other Logical Essays. London:
  Routledge and Kegan Paul Ltd; 1931. p. 156--198.

\bibitem{Popper1959}
Popper KR.
\newblock The logic of scientific discovery.
\newblock London: Hutchinson \& Co.; 1959.

\bibitem{Kyburg1964}
Kyburg HE, Smokier HE, editors.
\newblock Studies in subjective probability.
\newblock New York, NY: Wiley; 1964.

\bibitem{Schmidt2009}
Schmidt S.
\newblock Shall we really do it again? The powerful concept of replication is
  neglected in the social sciences.
\newblock Review of General Psychology. 2009;13(2):90--100.
\newblock doi:{10.1037/a0015108}.

\bibitem{Ioannidis2005}
Ioannidis JPA.
\newblock Why most published research findings are false.
\newblock {PLOS} Medicine. 2005;2(8):e124.
\newblock doi:{10.1371/journal.pmed.0020124}.

\bibitem{McElreath2015}
McElreath R, Smaldino PE.
\newblock Replication, communication, and the population dynamics of scientific
  discovery.
\newblock {PLOS} {ONE}. 2015;10(8):e0136088.
\newblock doi:{10.1371/journal.pone.0136088}.

\bibitem{Smaldino2016}
Smaldino PE, McElreath R.
\newblock The natural selection of bad science.
\newblock Royal Society Open Science. 2016;3(9):160384.
\newblock doi:{10.1098/rsos.160384}.

\bibitem{Higginson2016}
Higginson AD, Munaf{\`o} MR.
\newblock Current incentives for scientists lead to underpowered studies with
  erroneous conclusions.
\newblock {PLOS} Biology. 2016;14(11):e2000995.
\newblock doi:{10.1371/journal.pbio.2000995}.

\bibitem{Nissen2016}
Nissen SB, Magidson T, Gross K, Bergstrom CT.
\newblock Publication bias and the canonization of false facts.
\newblock {eLife}. 2016;5:e21451.
\newblock doi:{10.7554/elife.21451}.

\bibitem{Gelman2017}
Gelman A, Carlin J.
\newblock Some natural solutions to the p-value communication problem—and why
  they won't work.
\newblock Journal of the American Statistical Association.
  2017;112(519):899--901.
\newblock doi:{10.1080/01621459.2017.1311263}.

\bibitem{Bensonetal2017}
Benson AR, Gleich DF, Lim LH.
\newblock The spacey random walk: A stochastic process for higher-order data.
\newblock {SIAM} Review. 2017;59(2):321--345.
\newblock doi:{10.1137/16m1074023}.

\bibitem{Epstein2006}
Epstein JM.
\newblock Generative social science: Studies in agent-based computational
  modeling.
\newblock Princeton University Press; 2006.

\bibitem{Gilbert2008}
Gilbert N.
\newblock Agent-based models. vol. 153.
\newblock {SAGE} Publications, Inc; 2008.

\bibitem{Weisberg2009}
Weisberg M, Muldoon R.
\newblock Epistemic landscapes and the division of cognitive labor.
\newblock Philosophy of Science. 2009;76(2):225--252.
\newblock doi:{10.1086/644786}.

\bibitem{Muldoon2013}
Muldoon R.
\newblock Diversity and the division of cognitive labor.
\newblock Philosophy Compass. 2013;8(2):117--125.
\newblock doi:{10.1111/phc3.12000}.

\bibitem{Alexander2015}
Alexander JM, Himmelreich J, Thompson C.
\newblock Epistemic landscapes, optimal search, and the division of cognitive
  labor.
\newblock Philosophy of Science. 2015;82(3):424--453.
\newblock doi:{10.1086/681766}.

\bibitem{Schwarz1978}
Schwarz G.
\newblock Estimating the dimension of a model.
\newblock The Annals of Statistics. 1978;6(2):461--464.

\bibitem{Baumgaertner2018}
\newblock Baumgaertner B, Devezer B, Buzbas EO, Nardin LG.
\newblock A model-centric analysis of openness, replication, and reproducibility,
\newblock ArXiv e-prints. 2018;Nov:arXiv:1811.04525.

\bibitem{Lindley2000}
Lindley DV.
\newblock Philosophy of Statistics.
\newblock Journal of the Royal Statistical Society: Series D (The Statistician). 2000;49(3):293--337.

\bibitem{Akaike1973}
Akaike H.
\newblock Information theory and an extension of the maximum likelihood
  principle.
\newblock In: Petrov BN, Csaki F, editors. Proceedings of the 2nd International
  Symposium on Information Theory. Budapest: Akademiai Kiado; 1973. p.
  267--281.

\bibitem{Akaike1974}
Akaike H.
\newblock A new look at the statistical model identification.
\newblock {IEEE} Transactions on Automatic Control. 1974;19(6):716--723.
\newblock doi:{10.1109/tac.1974.1100705}.

\bibitem{Mule2017}
Gonzalez-Mul{\'e} E, Aguinis H.
\newblock Advancing theory by assessing boundary conditions with
  metaregression: A critical review and best-practice recommendations.
\newblock Journal of Management. 2017;44(6):2246--2273.
\newblock doi:{10.1177/0149206317710723}.

\bibitem{Whetten1989}
Whetten DA.
\newblock What constitutes a theoretical contribution?
\newblock Academy of Management Review. 1989;14(4):490--495.

\bibitem{Lv2013}
Lv J, Liu JS.
\newblock Model selection principles in misspecified models.
\newblock Journal of the Royal Statistical Society: Series B (Statistical
  Methodology). 2013;76(1):141--167.
\newblock doi:{10.1111/rssb.12023}.

\bibitem{Shiffrin2018}
Shiffrin RM, B{\"o}rner K, Stigler SM.
\newblock Scientific progress despite irreproducibility: A seeming paradox.
\newblock Proceedings of the National Academy of Sciences.
  2018;115(11):2632--2639.
\newblock doi:{10.1073/pnas.1711786114}.

\bibitem{Zollman2009}
Zollman KJS.
\newblock The epistemic benefit of transient diversity.
\newblock Erkenntnis. 2009;72(1):17--35.
\newblock doi:{10.1007/s10670-009-9194-6}.

\bibitem{Muthukrishna2019}
Muthukrishna M, Henrich J.
\newblock A problem in theory.
\newblock Nature Human Behaviour. 2019;
\newblock https://doi.org/10.1038/s41562-018-0522-1

\bibitem{Box1976}
Box GEP.
\newblock Science and statistics.
\newblock Journal of the American Statistical Association. 1976;71(356):791--799.

\bibitem{Nelder1986}
Nelder JA.
\newblock Statistics, science and technology.
\newblock Journal of the Royal Statistical Society: Series A (General). 1986;149(2):109--121.


\end{thebibliography}
\end{document}